# Orthorhombic nitride perovskite $CeTaN_{3-\delta}$ with switchable and robust ferroelectric polarization

Guozhu Song[1,7,†], Xiangliang Zheng[1,†], Xiaodong Yao[1], Xuefeng Zhou[1], Chao Gu[1,2], Qinghua Zhang,[3] Jian Chen[1], Chenglu Huang[1], Tiancheng Yang[1], Leiming Fang[4], Ping Miao[5,6], Lingxiang Bao[5,6], Wen Yin[5,6], Xiaohui Yu[3]， Jinlong Zhu[1,2], Wei Bao[7], Yusheng Zhao[1,8], Er-Jia Guo[3,*], and Shanmin Wang[1,2,*]

[1]*Department of Physics and Guangdong Basic Research Center of Excellence for Quantum Science, Southern University of Science & Technology, Shenzhen, Guangdong, 518055, China*
[2]*Quantum Science Center of Guangdong-Hongkong-Macao Greater Bay Area, Shenzhen, Guangdong, 518045, China*
[3]*Beijing National Laboratory for Condensed Matter Physics and Institute of Physics, Chinese Academy of Sciences, Beijing 100190, China*
[4]*Key Laboratory for Neutron Physics, Institute of Nuclear Physics and Chemistry, China Academy of Engineering Physics, Mianyang, 621999, China*
[5]*Institute of High Energy Physics, Chinese Academy of Sciences, Beijing, 100049, China*
[6]*Spallation Neutron Source Science Center, Dongguan, Guangdong, 523803, China*
[7]*Department of Physics, City University of Hong Kong, Tat Chee Avenue, Kowloon, Hong Kong, China*
[8]*College of Science, Eastern Institute of Technology, Ningbo, Zhejiang, 315200, China*

**Abstract**

Perovskite-type ternary nitrides with predicted exciting ferroelectricity and many other outstanding properties hold great promise to be an emerging class of advanced ferroelectrics for manufacturing diverse technologically important devices. However, such nitride ferroelectrics have not yet been experimentally identified, mainly due to the challenging sample synthesis by traditional methods at ambient pressure. Here we report the successful high-pressure synthesis of a high-quality ferroelectric nitride perovskite of $CeTaN_{3-\delta}$ with nitrogen deficiency, adopting an orthorhombic $Pmn2_1$ polar structure. This material is electrically insulating and exhibits switchable and robust electric polarization for producing ferroelectricity. Furthermore, a number of other extraordinary properties are also revealed in this nitride such as excellent mechanical properties and chemical inertness, which would make it practically useful for many device-relevant applications and fundamentally important for the study of condensed-matter physics.

***Keywords:*** *nitride perovskite, $CeTaN_{3-\delta}$, high pressure synthesis, ferroelectricity*

The quest for oxygen-free nitride ferroelectrics with robust polarization is crucially important for both the fundamental studies and practical applications in many fields of science and technology[1,2]. In particular, those nitride materials are promising for fabricating many important devices such as non-volatile memories[1-5], because they can largely mitigate disadvantages of its oxide counterparts (e.g., $Pb(Zr,Ti)O_3$[6]) that usually involve toxic elements and are technically challenging to integrate with semiconductors[1,4]. Generally, materials with ferroelectricity are electrically insulating, which, however, have been primarily limited to oxides and halides[7].

Nitrogen with one electron less than oxygen often favors forming metallic binary transition-metal (TM) nitrides due to its reduced electronegativity[8,9]. According to recent calculations[8], by adding alkali/alkaline- or rare-earth metals that play as electron donors, the covalent TM-N bonding can be substantially promoted to produce stable nitrogen-rich ternary nitrides. Of particular interest is the nitride perovskites $ABN_3$ with A = rare-earth metals and B = late TMs[10], and some of them have been predicted to be ferroelectrics [11-15], because of the TM-N hybridization that weakens short-range repulsions for driving polar distortion of $\{BN_6\}$ octahedra to open a bandgap, favoring a long-range ferroelectric order[16,17]. Besides, the thus-formed strong covalent metal-nitrogen bonds of these nitrides will endow them with excellent properties regarding their mechanical stiffness, chemical inertness, structural stability, and defect tolerance[8,9]. These, along with the ease of integration with the existing GaN-based electronics, would enable them to be an exciting class of versatile ferroelectrics for advancing the associated micro- and nano-electronics technologies[1,4]. However, due to the difficulties in preparing these nitrides by traditional methods at ambient pressure, the predicted ferroelectricity of nitride perovskites has yet to be clearly revealed by experiments.

In fact, most the nitride-perovskite products of traditional methods are actually oxynitrides (e.g., $LaWO_{0.6}N_{2.4}$)[18], excepting a less-explored case of $ThTaN_3$ that was previously reported[19]. The use of physical vapor deposition has recently led to the formation of thin-film $LaWN_{3-\delta}$[2] and $CeBN_3$ (B = W or Mo)[20], but no convincible ferroelectricity was determined due to the significant current leakage[2]. On a note, crystallinities of those reported samples are too low to obtain reliable diffraction data for accurate structure resolution, hence raising many contentious issues regarding their crystal structures[17].

High-pressure (P) synthesis is a proven powerful approach for preparation of TM nitrides, because pressure can largely promote the chemical activities of both TM and N atoms for forming nitrogen-rich nitrides[8,21-26]. Recently, we have formulated innovative high-P reaction routes for successfully preparing high-quality $LaWN_{3-\delta}$ samples[17], based on which its crystal structure and many other properties have been well explored. Note that a high-P synthesis of $LaReN_3$ was also reported[27]. However, those two nitrides are either a narrow-bandgap semiconductor or intrinsic metal, incapable of producing switchable ferroelectric polarization. Besides, electronic properties of such perovskites tend to strongly depend on the valence-electron number of B-site atom. In this regard, substituting W by Ta is anticipated to form Ta-bearing nitride perovskites with enhanced electrical insulation, due to the reduced valence electrons in each unit cell of the resultant nitrides. Recent calculations have indeed predicted a largely-gapped candidate of $CeTaN_3$ exhibiting favorable ferroelectricity[15,28], but it has not yet been practically identified, calling for rigorous experimental investigations.

Here we extend the formulated high-P methods to the Ce-Ta-N ternary system, leading to the attainment of high-quality $CeTaN_{3-\delta}$ bulk samples. Our comprehensive investigations reveal that the nitride is a nitrogen-deficient $CeTaN_{2.68}$ with an orthorhombic $Pmn2_1$ symmetry (see ref. [29] for Experimental details), showing extraordinary ferroelectric performances and other superior properties.

**Results and Discussion**

Based on the compositional and structural characterizations of the final samples synthesized through reactions involving $NaNH_2$ with CeN and Ta or $CeO_2$ and $Ta_2O_5$ at 5 GPa and temperatures of 1475-1873 K (Fig. 1 and Figs. S1-S2), a perovskite phase of $CeTaN_3$ is obtained. The time-of-flight secondary ion mass spectrometry (ToF-SIMS) measurements of a single-crystal sample show that the concentrations of Na, O, and other impurities are well below the detection levels. As judged from the tolerance factor of $t$<1 [29-31], the nitride should adopt either a tetragonal or orthorhombic symmetry. Accordingly, its structural candidates mainly include orthorhombic symmetries of $Pmc2_1$, $Pna2_1$, $Pmn2_1$, and *Pnma* and tetragonal ones of *P*4*mbm* and *I*4*mcm*[15,28,30]. By analysis of XRD and NPD data (Figs. S3-S6), *Pnma*, and tetragonal structures can be quickly excluded. Therefore, $Pmn2_1$, $Pmc2_1$, and $Pna2_1$ with broken inversion symmetries are the most probable candidates, but they reach almost the same excellence of refinements based on both XRD and NPD data (Figs. 1a-1b, Figs. S7-S8, and Table S1), making them hardly distinguishable. Note that the nitrogen is an excellent neutron scatter, allowing accurately determining both the atomic positions and deficiencies of N atoms[32], different from x-rays that is nearly transparent to nitrogen.

Electron diffraction along certain crystallographic directions enables collecting strong diffraction signals for the otherwise weak reflections, by which the structures can be differentiated (Fig. S9). We performed TEM observations along five distinct zone axes and used SAED measurements to identify the reflections and definitively resolve the crystal structure (Figs. 1c-1d and Fig S10). By tilting the zone axes (Fig. S11), we confirmed that all observed diffraction spots arise solely from the primary diffraction [33]. According to our simulations, the $Pmc2_1$ model gives rise to additional diffraction spots (e.g., the 100 spots) and systematic extinctions of certain rows of reflections (e.g., $h0\bar{1}$ and $h0\bar{3}$) along specific zone axes (Fig. S10). These features are not observed in the experimental data, allowing $Pmc2_1$ to be readily ruled out. Using $Pna2_1$, the similar extinction of diffraction-spot rows is also observed (Fig. S10). However, this model predicts a zero displacement of Ta atom relative to the center of its surrounding nitrogen octahedral coordination along the $[010]_{Pna2_1}$ direction (Fig. S12), which is inconsistent with the atomic positions determined from iDPC observations. Therefore, $Pna2_1$ can also be clearly excluded. In contrast, all the diffraction spots can be excellently reproduced by the $Pmn2_1$ model across all the involved five zone axes (Figs. 1e-1f and Fig. S10), implying it is the most appropriate symmetry for $CeTaN_{3-\delta}$, as opposed to $Pmc2_1$ and $Pna2_1$. This conclusion is further supported by the excellent agreements between $Pmn2_1$ structures and atomic-scale iDPC- and ABF-STEM images (Figs. S12-S14). However, our refinements indicate that the disordered nitrogen deficiency, δ, is involved in $Pmn2_1$-$CeTaN_{3-\delta}$, with a refined composition of $CeTaN_{2.68}$ (i.e., δ≈0.32).

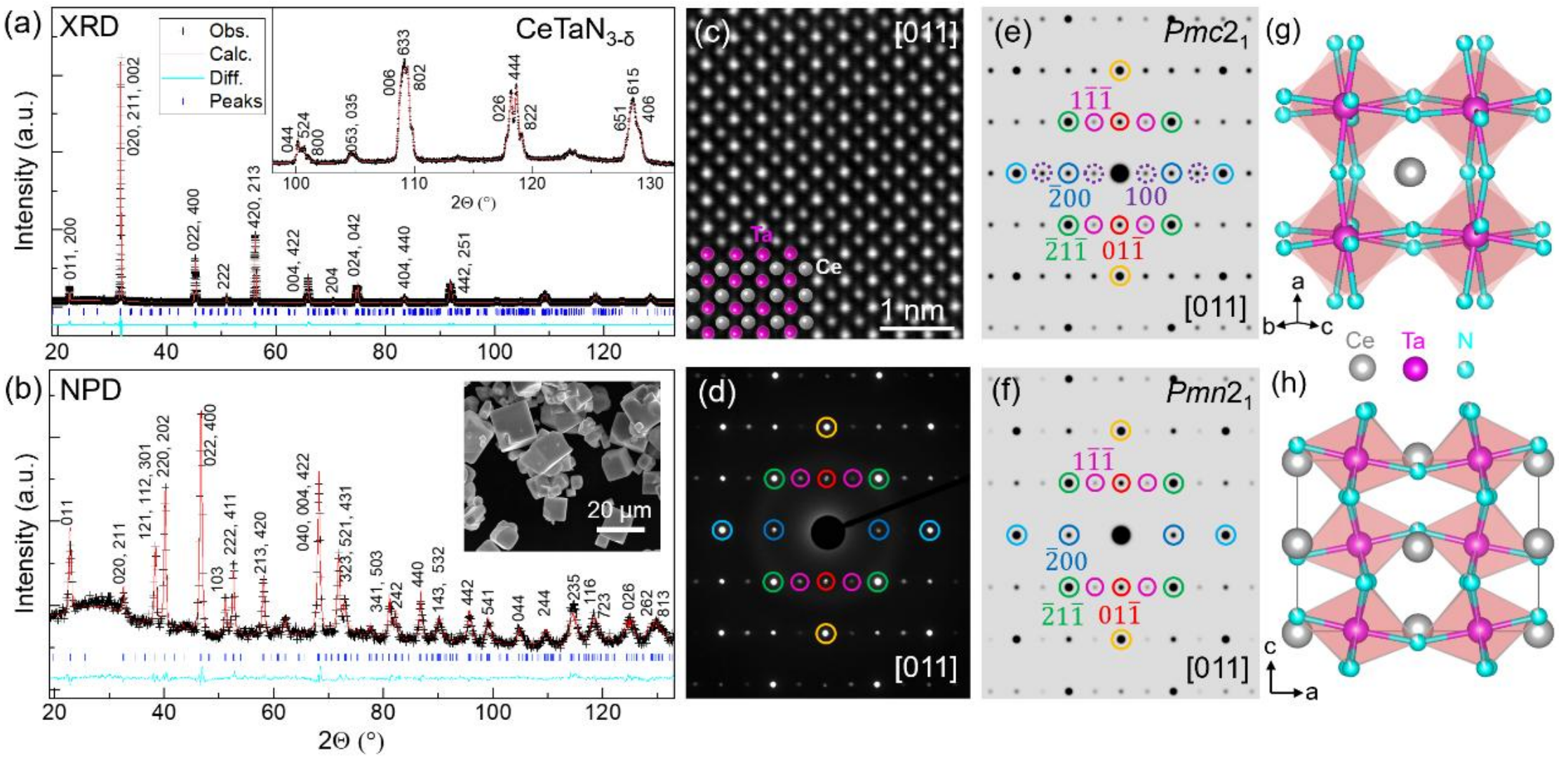

**FIG. 1. Structure resolution for $CeTaN_{3-\delta}$ synthesized at 5 GPa and 1873 K.** (**a**)-(**b**) Refined ambient x-ray diffraction (XRD) and neutron powder diffraction (NPD) patterns, using the $Pmn2_1$ symmetry [46]. (**c**) High-angle annular dark-field scanning transmission electron microscopy (HAADF-STEM) observation. (**d**) Selected area electron diffraction (SAED) pattern. (**e**)-(**f**) Simulated SAED patterns using $Pmc2_1$ and $Pmn2_1$. (**g**)-(**h**) Structures of $Pmn2_1$-$CeTaN_{3-\delta}$.

Apparently, the cationic valance states of $CeTaN_{2.68}$ are nominally $Ta^{5+}$ and $Ce^{3+}$, consistent with the determined by XPS experiments (Fig. S2) [34,35]. Compared with $Ce^{4+}$ in its ideal stoichiometric $CeTaN_3$, $Ce^{3+}$ has a larger ionic radius and the resultant tolerance factor is as large as $t\approx0.94$ [29], which is favorable for stabilizing a perovskite-type structure, rationalizing its substoichiometry. The thus-refined lattice parameters of $Pmn2_1$-$CeTaN_{3-\delta}$ are listed in Table 1, compared to those of the two excluded $Pna2_1$ and $Pmc2_1$-$CeTaN_{3-\delta}$ (Tables S2-S3). Solving such a challenging crystal structure of $CeTaN_{3-\delta}$ is critically important for study of its foundational properties. The presence of atomic deficiency in this nitride can largely alter its structural stability and many intrinsic properties, which inevitably poses insurmountable challenges for accurate predictions [15].

Compared with its pristine $Pm\bar{3}m$ phase, $Pmn2_1$ is associated with antiphase and in-phase antiferrodistortive instabilities [17,36], characterized by rotating the neighboring $\{TaN_6\}$ octahedra with two distortion angles of ~157 and 158° (Fig. S12). The broken inversion symmetry of $Pmn2_1$ phase would lead to an effective electric dipole in each unit cell, which is often active for the nonlinear second harmonic generation (SHG) measurements. To check the detailed rotational symmetry of $CeTaN_{3-\delta}$, angle-resolved SHG measurements are also conducted (Fig. S15). Both the *s*-out (i.e., $I_{\perp}$) and *p*-out (i.e., $I_{\parallel}$) signals exhibit a fourfold rotational symmetry, consistent with that of the *mm*2 point group. The domain effect is also identified by nonzero background signals [37]. Polar distortions of Ta relative to its surrounding N coordinates in either $Pmn2_1$, $Pmc2_1$ or $Pna2_1$ can effectively relieve the filling of the Ta:5d-N:2p antibonding states for opening a bandgap, as reported in $LaWN_3$ [17]. By optical absorption experiments, a large bandgap of $E_g$ = 1.5 eV is determined for $CeTaN_{3-\delta}$ (Fig. S15), in stark contrast to that of $LaWN_{3-\delta}$ with a small one (i.e., 0.66 eV). Electrical transport measurements show that $CeTaN_{3-\delta}$ has infinitely high resistivities, a characteristic of an insulator. No long-range magnetic ordering is observed at low temperatures (Fig. S17), indicative of a paramagnetic nitride. Because of the presence of strong metal-nitrogen bonds, this nitride has a superior hardness of ~6 GPa over most oxide perovskites (Fig. S18) [38].

**Table 1. Refined lattice parameters of $Pmn2_1$-$CeTaN_{3-\delta}$ based on both XRD and NPD data at ambient conditions.**

| $Pmn2_1$-$CeTaN_{3-\delta}$ | | |
|---|---|---|
| Formula | $CeTaN_{2.68}$ | |
| Symmetry | Orthorhombic, $Pmn2_1$ (No. 31) | |
| Cell parameter (Å) | $a$ = 8.0064 (2), $b$ = 5.7077 (3), $c$ = 5.6825 (4) | |
| Cell volume (Å$^3$) | 259.679 | |
| **Atomic position** | **Wyckoff site** | **Occupancy** |
| | Ce: 2*a* (1/2, 0.246, 0.013) | 1 |
| | Ce: 2*b* (1.0, 0.252, 0.999) | 1 |
| | Ta: 4*c* (0.257, 0.251, 0.506) | 1 |
| | N1: 4*c* (0.293, 0.508, 0.289) | 0.91 |
| | N2: 4*c* (0.222, 0.010, 0.785) | 0.92 |
| | N3: 2*a* (1/2, 0.763, 0.936) | 0.85 |
| | N4: 2*b* (1.0, 0.745, 0.087) | 0.86 |
| *wRp* (%) | 3.5 for XRD, 4.5 for NPD | |

By examining structure of $Pmn2_1$-$CeTaN_{3-\delta}$, the spontaneous polarization can be identified. Along the *a*- or *b*-axis, two adjacent N atomic arrays are symmetrically distributed around their center Ta array with an equal Ta-N distance of $\frac{1}{4}a\approx2$ Å and $\frac{1}{4}b\approx1.43$ Å, respectively (Figs. 2a-2b), without involving net charge polarizations. In contrast, along the *c*-axis there is an obvious shift of the Ta layer relative to the center line of two neighboring N layers (Fig. 2c), which gives rise to two inequal Ta-N distances of $0.28c\approx1.59$ Å and $0.22c\approx1.25$ Å, respectively, for separating the positive and negative charge centers, close to the observed by iDPC-STEM measurements. As a result, a net electric polarization is produced along this direction. According to the point-charge model [39-42], the material's theoretical polarization ($P_{theo.}$) can be determined by calculation of the effective dipole moments per unit-cell volume (see Eq. S3 [29]), and the thus-obtained value is $P_{theo.}$= 29.9 μC/cm$^2$, comparable to or even larger than that of typical oxide ferroelectrics as exemplified by $BaTiO_3$ (i.e., 26 μC/cm$^2$) [43].

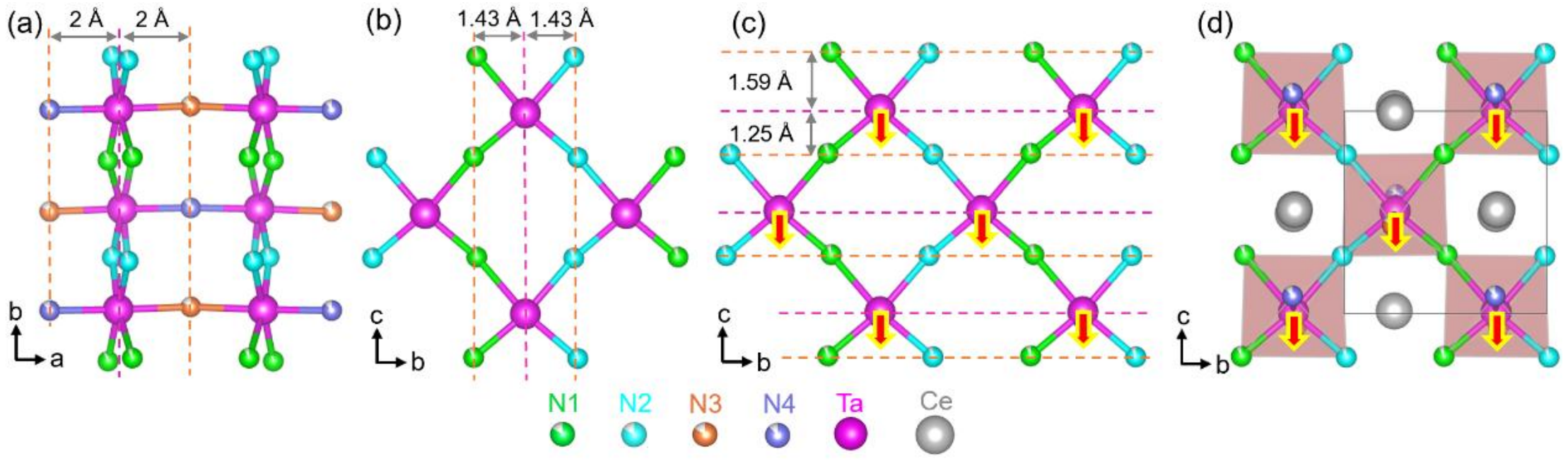

**FIG. 2. Identification of electric polarization and polar symmetry in *Pmn*2₁-CeTaN$_{3\text{-}\delta}$. (a)-(c)** Analysis of polarizations along axial directions. (**d**) Crystal structure with a nonzero polarization along the *c*-axis. Arrows in (c)-(d) denote relative displacements of Ta relative to its surrounding N.

To determine ferroelectric properties of single-crystal CeTaN$_{3\text{-}\delta}$ with a small crystallite size of 30-40 μm, special electrodes are designed based on a diamond-anvil cell (Fig. 3a), leading to successful observations of strong ferroelectric responses (Figs. 3b-3c). By sweeping the electric field ($E$) between peak values of $E_m$ = ±5 kV/cm, the well-defined hysteresis loops are obtained for both the polarization ($P$) and current density ($J$). In particular, two sharp current peaks appear at ~3 kV/cm (Fig. 3b), corresponding to the flipping of a long-range ferroelectric order (i.e., switchable polarization). Given these phenomena, the room-temperature ferroelectricity of CeTaN$_{3\text{-}\delta}$ is thus unambiguously revealed with a stable polarization. Moreover, its coercive field ($E_c$) is determined to be ~3 kV/cm, smaller than that of many bulk oxide ferroelectrics (e.g., 8~10 kV/cm for $BaTiO_3$ [44]). A small $E_c$ should be favorable for fabricating high-precision devices for low-voltage circuits.

We note that the observed P-E hysteresis loop of CeTaN$_{3\text{-}\delta}$ slightly deviates from the square-like shape [4,45], similar to the observed in many oxide ferroelectrics [45]. By varying $E_m$ from 2.3 kV/cm to its dielectric breakdown limit of 6.3 kV/cm, an abrupt loop broadening is occurred around $E_c$ = 3 kV/cm, as the polarization starts to be turned over. The subsequent increase of $E_m$ leads to a rapid increase of both the remnant ($P_r$) and saturated ($P_s$) polarizations, eventually achieving asymptotic values of 18.5 and 20.5 μC/cm$^2$, respectively, which are smaller than the theoretical value (i.e., 29.9 μC/cm$^2$). We suspect that the unavoidable deficiencies such as nitrogen vacancies or cation nonstoichiometry can significantly impact both electrical conductivity and electric polarization. The vacancies often increase carrier concentration and conductivity, while also inducing internal fields that pin domains and alter polarization. On the other hand, the cation imbalances may modify lattice symmetry and electronic structure, thereby altering ferroelectric responses accordingly. It is worthwhile to mention that our work is the first demonstration of a nitride perovskite with room-temperature ferroelectricity and a large stable polarization, overcoming longstanding challenges in synthesizing nitride ferroelectrics.

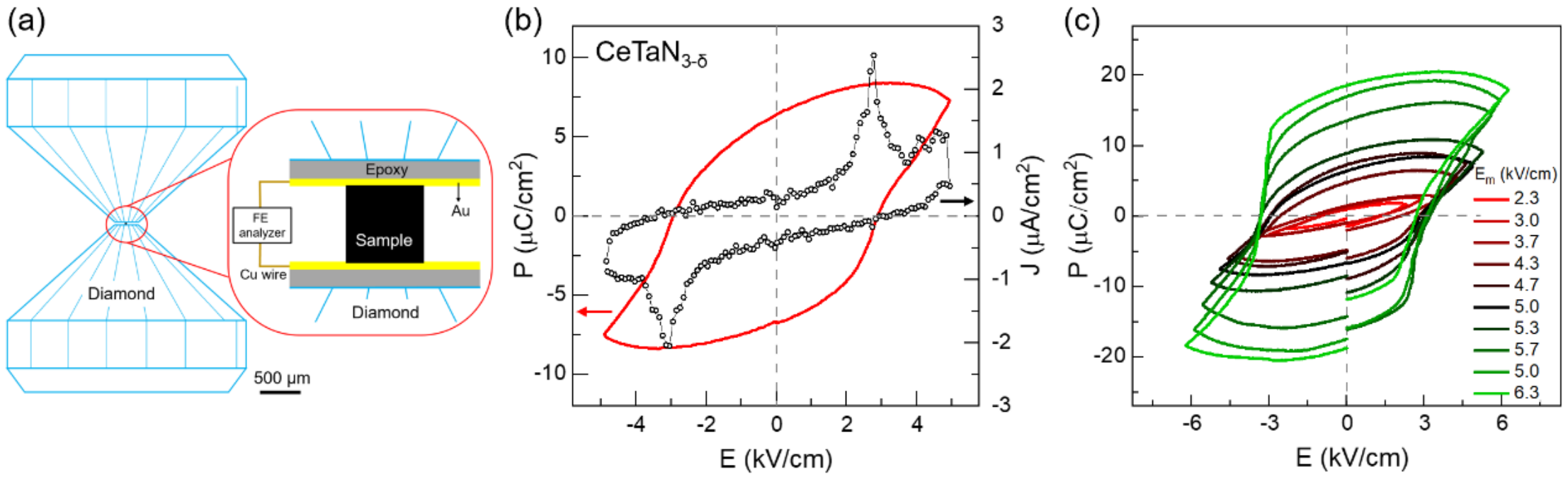

**FIG. 3. Ferroelectric measurements of single-crystal $CeTaN_{3-\delta}$ at ambient conditions.** (**a**) Experimental setup for ferroelectric (FE) measurements. (**b**) P-E and J-E hysteresis loops measured at 500 Hz by sweeping the electric field within -5 to +5 kV/cm. (**c**) Cyclic experiments for measuring P-E hysteresis loops at 500 HZ and various peak electric-field values of $E_m$.

Variable-T XRD and NPD and high-T dilatometric measurements show that $Pmn2_1$-$CeTaN_{3-\delta}$ possesses superior thermal stabilities in the 4-1600 K temperature range in vacuum without involving any phase transition (Fig. 4a and Figs. S19-S21), consistent with TGA experiments in argon (Fig. 4b). Above 1600 K, $CeTaN_{3-\delta}$ starts to decompose into binary metal nitrides (Fig. S21). In air, the sample possesses a high oxidization temperature of 727 K. Our further experiments also show that $CeTaN_{3-\delta}$ exhibits excellent chemical inertness for resisting both acidic and alkaline conditions, indicating that it is suitable for applications in hostile environments.

By analysis of high-T NPD data (Table S4), the displacement of positive charge center relative to the negative charge center ($\Delta d$) can be deduced (Fig. S16), and $P_{\text{theo.}}$ can be easily calculated (Fig. 4c). Both $\Delta d$ and $P_{\text{theo.}}$ have a similar trend of monotonic decrease with increasing temperature. The spontaneous polarization can persist up to 1373 K, at which a large polarization of $P_{\text{theo.}} \approx 17.8\ \mu C/cm^2$ is obtained, in stark contrast to most the oxide ferroelectrics with fragile polarizations above 700 K. Besides, our further experiment shows that the dielectric constant of this material remains nearly invariant up to 800 K (Fig. S21). Unlike oxide counterparts with fragile polarizations that are dominated by ionicity, the remarkable robustness of polarization in $CeTaN_{3-\delta}$ is presumably attributed to its strong metal-nitrogen bonding [8]. Hence, its polarization switching dynamics and domain-wall mobility may also be profoundly changed for producing exotic phenomena. Besides, the interplay of covalency and polarization in nitrides would challenge the traditional soft-mode mechanism of ferroelectricity for advancing the associated theories.

Apparently, the intriguing combination of outstanding ferroelectricity and many other properties in this nitride would make it highly promising for a broad range of innovative device-relevant applications. The discovery of the remarkable nitride ferroelectric will not only offer a versatile platform for investigating ferroelectricity but also encourage further research into ternary transition-metal nitrides. This largely unexplored chemical space holds immense potential for uncovering exciting materials and studying various exotic phenomena at the frontiers of material science and condensed-matter physics [8,9].

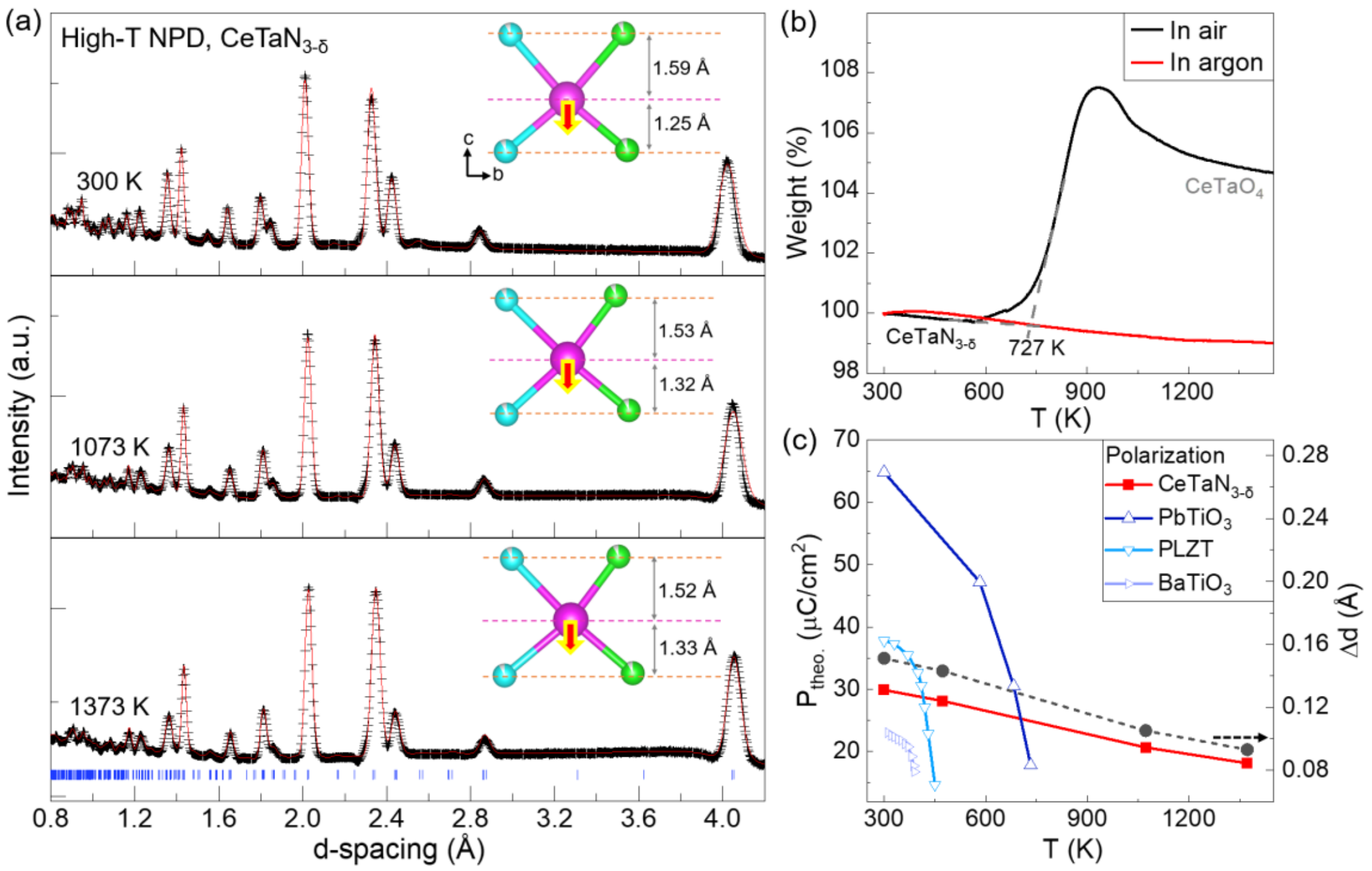

**FIG. 4. Phase stability and ferroelectric polarization of *Pmn*2₁-$CeTaN_{3-\delta}$ at high temperatures.** (**a**) Refined high-T NPD patterns collected at different temperatures of 300, 1073, and 1373 K. Insets are the associated {$TaN_6$} octahedral units. (**b**) Thermogravimetric analysis (TGA) measurements in air and argon. (**c**) Temperature dependence of $P_{theo.}$ and Δd obtained based on the point-charge model[39]. The cases of typical oxide ferroelectrics are also added for comparison including $PbTiO_3$[47], (Pb,La)(Zr,Ti)$O_3$ (i.e., PLZT)[48], and $BaTiO_3$[49].

## Conclusions

In summary, using high-P reaction routes, we have successfully synthesized high-quality nitrogen-deficient $CeTaN_{3-\delta}$ bulk samples, leading to resolution of its crystal structure with a $Pmn2_1$ polar symmetry and a composition of $CeTaN_{2.68}$. The material is an insulator and exhibits a switchable ferroelectric polarization along the *c*-axis with a considerably large electric polarization of ~20.5 μC/cm$^2$ and a suitable coercive field of ~3 kV/cm. Besides, the polar state of $CeTaN_{3-\delta}$ can be sustained up to 1373 K, representing the most robust electric polarization of materials among the known ferroelectrics.

## Acknowledgments

We acknowledge the financial support from the National Key R&D Program of China (Grant Nos. 2021YFA1400300 and 2020YFA0309100), the National Natural Science Foundation of China (Grant Nos. 12474013, 12174175, and U22A20263), the Guangdong Basic and Applied Basic Research Foundation (Grant Nos. 2022B1515120014, 2023B0303000003, and 2023B1515120060), the Guangdong Provincial Quantum Science Strategic Initiative (Grant No. GDZX2201001), the Shenzhen Basic Research Fund (Grant No. JCYJ20220530113016038), the CAS Project for Young Scientists in Basic Research (Grant No. YSBR-084), and the Beijing Natural Science Foundation (Grant No. JQ24002).

*Supplemental Material for*

# Orthorhombic nitride perovskite $CeTaN_{3-\delta}$ with switchable and robust ferroelectric polarization

Guozhu Song[1,7,†], Xiangliang Zheng[1,†], Xiaodong Yao[1], Xuefeng Zhou[1], Chao Gu[1,2], Qinghua Zhang,[3] Jian Chen[1], Chenglu Huang[1], Tiancheng Yang[1], Leiming Fang[4], Ping Miao[5,6], Lingxiang Bao[5,6], Wen Yin[5,6], Xiaohui Yu[3], Jinlong Zhu[1,2], Wei Bao[7], Yusheng Zhao[1,8], Er-Jia Guo[3,*], and Shanmin Wang[1,2,*]

[1]Department of Physics and Guangdong Basic Research Center of Excellence for Quantum Science, Southern University of Science & Technology, Shenzhen, Guangdong, 518055, China

[2]Quantum Science Center of Guangdong-Hongkong-Macao Greater Bay Area, Shenzhen, Guangdong, 518045, China

[3]Beijing National Laboratory for Condensed Matter Physics and Institute of Physics, Chinese Academy of Sciences, Beijing 100190, China

[4]Key Laboratory for Neutron Physics, Institute of Nuclear Physics and Chemistry, China Academy of Engineering Physics, Mianyang, 621999, China

[5]Institute of High Energy Physics, Chinese Academy of Sciences, Beijing, 100049, China

[6]Spallation Neutron Source Science Center, Dongguan, Guangdong, 523803, China

[7]Department of Physics, City University of Hong Kong, Tat Chee Avenue, Kowloon, Hong Kong, China

[8]College of Science, Eastern Institute of Technology, Ningbo, Zhejiang, 315200, China

**Email**: wangsm@sustech.edu.cn (S. Wang); ejguo@iphy.ac.cn (E. J. Guo).

## Experimental details

***Synthesis.*** Two reaction routes were used for synthesizing $CeTaN_{3-\delta}$ under high pressure (P) and high temperature (T) conditions, similar to those of $LaWN_{3-\delta}$[1]. The first reaction route is an oxygen-free process through the reaction between CeN (>99.5%, Macklin), Ta (>99.5%, Macklin), and $NaNH_2$ (>99.9%, Alfa Aesar), without involving any oxide precursor. Because the byproducts of this route are either a gas or chemically highly unstable compound in air, it is really difficult to identify the associated phases. Nevertheless, a reaction equation can be tentatively proposed as following,

$$CeN + Ta + 2NaNH_2 = CeTaN_3 + 2NaH + H_2 \qquad \text{(S1)}.$$

The second reaction route involves the nitridation of starting oxides of $CeO_2$ (>99.5%, Aladdin) and $Ta_2O_5$ (>99.9%, Alfa Aesar), using $NaNH_2$ as a nitrogen source, given by,

$$2CeO_2 + Ta_2O_5 + 9NaNH_2 = 2CeTaN_3 + 9NaOH + 3NH_3 \qquad \text{(S2)}.$$

The involved byproduct of NaOH was identified by the XRD measurement. A sharp smell of urine-like odor was released from the recovered sample upon opening the press, suggesting the formation of $NH_3$ gas. In each the above-mentioned reaction process, excess $NaNH_2$ was used to eliminate unwanted precursors in the final product and maintain a nitrogen-rich environment for forming nitrogen-rich nitrides. The starting-reactant molar ratios of CeN: Ta: $NaNH_2$ = 1: 1: 10 and $CeO_2$: $Ta_2O_5$: $NaNH_2$ = 2: 1: 10 were used for these two reaction routes, respectively. Referring to ref. [1] for more details.

High P-T synthesis experiments were carried out by using a DS 6 × 10 MN cubic press installed in the high-P lab of SUSTech[2]. Prior to the experiment, the homogenously mixed powers of the related starting reactants were compacted into a cylindrical pellet of 12 mm in diameter and 10 mm in height and then encapsulated by an *h*BN or Mo capsule before loading into the pre-prepared high-P cell assembly. Because of the sensitivity of $NaNH_2$ to air, all those procedures were done in an argon-filled glovebox. In each experimental run, we first compressed the cell to 5 GPa and subsequently heated the cell to a target temperature of 1073-1873 K and soaked for 20 min before quenching by directly shutting off the heating power. The single-crystal samples of 10-50 μm in crystallite size were grown at 5 GPa and 1873 K for 20 min and were used for ferroelectric measurements. More experimental details can be found elsewhere[2]. In the recovered sample， the thus-formed small nitride crystals were uniformly distributed in the unreacted excess $NaNH_2$ and byproducts (e.g., NaOH or NaH). Because the byproducts and unreacted starting reactants are highly aqueous soluble, the nitride crystals can be readily purified by washing with distilled water, followed by drying in an oven at 353 K. To check the chemical inertness of the material, the as-purified sample powders were consecutively treated in aqueous solutions of HCl, $HNO_3$, and NaOH with weight percentages of 31%, 55%, and 20%, respectively, for 2 hours at room temperature; careful analysis of such-treated sample powders showed that they remained almost intact, strongly indicating that the material has excellent chemical inertness to effectively resist both acidic and alkaline conditions. Besides, in spite of a long-term exposure in air for several months, the sample did not involve any degradation such as the hygroscopic and aging phenomena.

***Physical property characterizations.*** The final products were checked by an x-ray diffractometer (XRD) with a Cu target. Low-T XRD experiment was carried out in the 4-300 K temperature range to study possible phase transitions and lattice thermal expansions at ambient pressure. Angle-dispersive neutron powder diffraction (NPD) measurement was performed at the neutron beamline of China Mianyang Research Reactor (CMRR), and the wavelength of incident neutron beam is λ = 1.5878 Å. Time-of-flight NPD data were collected at high temperatures to study structural evolutions, using the GPPD beamline of China Spallation Neutron Source (CSNS). Both the XRD and NPD data were analyzed using the GSAS program[3]. The simulations of XRD, NPD, and electron powder diffraction (EPD) patterns were performed by using the Reflex module of Materials Studio 5.0. Electron diffraction patterns were simulated by using the Crystal Maker. Scanning electron microscopy (SEM), energy-dispersive x-ray spectroscopy (EDS), and time-of-flight secondary ion mass spectrometry (ToF-SIMS) measurements were conducted to determine both the morphologies and chemical composition of as-synthesized samples. High-resolution transmission electron microscopy (HRTEM) experiment was done to observe the sample's microstructure at the atomic scale and study crystal structure of the nitride. The atomic-scale structures of samples were observed by selected area electron diffraction (SAED), high-angle annular dark-field (HAADF), annular bright-filed (ABF), and integrated differential phase contrast (iDPC) scanning transmission electron microscopy (STEM; Themis Z/USA)) experiments equipped with double spherical aberration (Cs) correctors. All the involved STEM experiments were performed with a same convergence semiangle of 24 mrad. The collection angles of 50 - 200 mrad and 12 - 24 mrad were used for the HAADF- and ABF-STEM measurements, respectively. The iDPC-STEM images were acquired with a 4-segmented annular detector and the phase reconstruction was automatically calculated by the instrument via the integrated center-of-mass (iCOM) algorithm.

Before experiments, the STEM specimens were prepared by standard mechanical thinning followed by ion milling. All TEM images were processed and analyzed using the Gatan DigitalMicrograph program. X-ray photoelectron spectroscopy (XPS) measurements were performed to explore the binding energies and electronic structures of the involved elements of samples.

Ultraviolet-visible (UV-Vis) absorption experiment was employed to measure the nitride's optical bandgap, based on single-crystal samples. DC magnetization data were collected using a SQUID VSM magnetometer (Quantum Design). Second-harmonic generation (SHG) measurements were conducted to check the inversion symmetry of samples, using a laser with an incident wavelength of $\lambda_0$ = 800 nm. Angle-dependent SHG measurements were also performed to investigate the rotational anisotropy of crystal structure, and the detailed experimental setup was described in Fig. S15. Thermogravimetric analysis (TG) experiments were carried out to investigate the sample's thermal stabilities in air and argon, respectively. Vickers hardness tests were performed based on a well-sintered polycrystalline bulk sample prepared at 10 GPa and 1473 K for 30 min, using a Kawaii-type multi-anvil large-volume apparatus. The sample's hardness of was then tested at different loads of 0.49, 0.98, 1.96, 4.9, and 9.8 N, respectively. At each load, the measurements were repeated for more than five times to obtain a statistic average. The temperature-dependent dielectric constant measurements were carried out in the 300 – 800 K temperature range at a frequency of 1000 Hz, using an impedance analyzer (KeysightE4980AL/USA). Prior to experiment, the as-synthesized sample powders of 0.5 g were tightly sandwiched by two waveguides of the network analyzer. Based on a well-sintered polycrystalline bulk sample (i.e., prepared at 12 GPa and 1173 K for 20 min), the coefficient of thermal expansion (CTE) of this nitride was also determined up to 1673 K in $N_2$ gas, by thermodilatometric measurements (Linseis L75V) with a heating rate of 2.5 K/min.

***Strategy for ferroelectric measurements***. Ferroelectric properties of the as-prepared nitride were determined by using a ferroelectric tester (the Precision Multiferroic-II, Radiant Technologies), based on single-crystal samples of 30-40 μm in crystallite size with cuboidal shapes. However, it is technically difficult to make electrodes for such small crystals using the traditional experimental setup of a typical ferroelectric measurement. Thus, a diamond-anvil cell (DAC) was used for solving this issue by a rational design of electrodes (see detailed setup in Fig. 4(a) of the main text). The culet size of diamond anvils is 300 μm. Different from the regular DAC experiments for the purpose of high-P study, in this work the DAC device was only used as a holder to tightly clamp the crystal, without involving a gasket system. Before loading the sample into the cell, the two opposite diamonds were uniformly covered with a thin layer of epoxy resin (i.e., ~100 μm in thickness). A gold electrode layer with a thickness of ~100 nm was then coated on top of the hardened epoxy resin layer of each diamond anvil, using a magnetron sputtering system (JGP450). Note that the epoxy resin layer is mechanically soft, which served as a cushioning substrate for protection of the sample crystal from being crashed. Besides, due to the presence of such an epoxy resin layer with excellent elasticity, electrical connections of the measurement circuit became more robust, allowing cyclic ferroelectric experiments based on the same one single crystal. In each experiment, a high-quality sample crystal was selected and subsequently loaded between the two as-prepared electrodes. During the measurement, by sweeping the applied voltage ($U$) at ambient conditions, the ferroelectric response signals of the sample were collected by a ferroelectric analyzer. The electric field ($E$) between the two opposite electrodes was calculated in terms of the applied voltage and sample's height ($h$), given by $E = U/h$.

***Evaluation of the tolerance factor.*** The Goldschmidt tolerance factor ($t$) of $ABN_3$ was calculated by $t = \frac{r_A + r_N}{\sqrt{2}(r_B + r_N)}$, where $r_A$, $r_B$, and $r_N$ are the effective ionic radii of the associated ions that have been well documented in ref. [4]. For the ideal stoichiometric $CeTaN_3$, the nominal ionic valence states are $Ce^{4+}$, $Ta^{5+}$, and $N^{3-}$ with $r_{Ce^{4+}}$= 1.14 Å, $r_{Ta^{5+}}$= 0.64 Å, and $r_{N^{3-}}$= 1.46 Å, respectively, giving rise to a very small tolerance factor of $t \approx 0.88$. Considering the presence of nitrogen deficiency, the actual valence state of Ce ion is nominally $Ce^{3+}$ according to our XPS measurements (see Fig. S2), so that the corresponding ionic radius is $r_{Ce^{3+}}$= 1.34 Å. In this case, the thus-deduced tolerance factor is $t \approx 0.94$, approaching the unity (i.e., $t = 1$) for its ideal pristine cubic perovskite. Because of $t < 1$ for $Pm\bar{3}m$-$CeTaN_3$, there must exist a profound bond-length mismatch for the buildup of compressive and tensile stresses in the Ta-N and Ce-N bonds, respectively. Therefore, crystal structure of $CeTaN_3$ tends to be distorted into either a tetragonal or orthorhombic symmetry by cooperative rotations of the corner-shared $\{TaN_6\}$ octahedra for relieving stresses. In fact, a number of structural candidates are available for this nitride including orthorhombic symmetries of $Pmn2_1$, $Pna2_1$, $Pmc2_1$, and $Pnma$ and tetragonal ones of $P4mbm$ and $I4mcm$, and they are the most frequently adopted structures for oxide perovskites[5-7].

Clearly, by reducing the Ce valence state from +4 to +3, the tolerance factor has a nearly 7% decrease to largely alleviate the bond-length mismatch between Ce-N and Ta-N bonds, which is favorable for structural stability. This partially explains the reason why the Ce ion has a realistic $Ce^{3+}$ valence state, instead of the ideal state of $Ce^{4+}$, accounting for its substoichiometry.

***Calculation of the electric polarization:*** Based on the well-refined crystal structure, the theoretical electric polarization ($P_{theo.}$) of the material was readily calculated based on the point-charge mode[8], given by

$$P_{theo.} = \frac{1}{V}\left(-e\frac{\sum(N_i\eta_i)Z_i}{\sum(N_i\eta_i)}\right) \quad \text{(S3).}$$

Where $V$ is the unit-cell volume, $e$ is the electron charge, $N_i$ is the ionic valence state (*i.e.,* +3, +5, and -3 for the involved ions of $Ce^{3+}$, $Ta^{5+}$, and $N^{3-}$, respectively), $\eta_i$ is the refined atomic occupancy, and $Z_i$ is the atomic position vector projected along the polarization direction (i.e., the *c*-axis for $Pmn2_1$-$CeTaN_{3-\delta}$). This model has extensively been used for evaluation of the polarization of many ferroelectric materials[9,10].

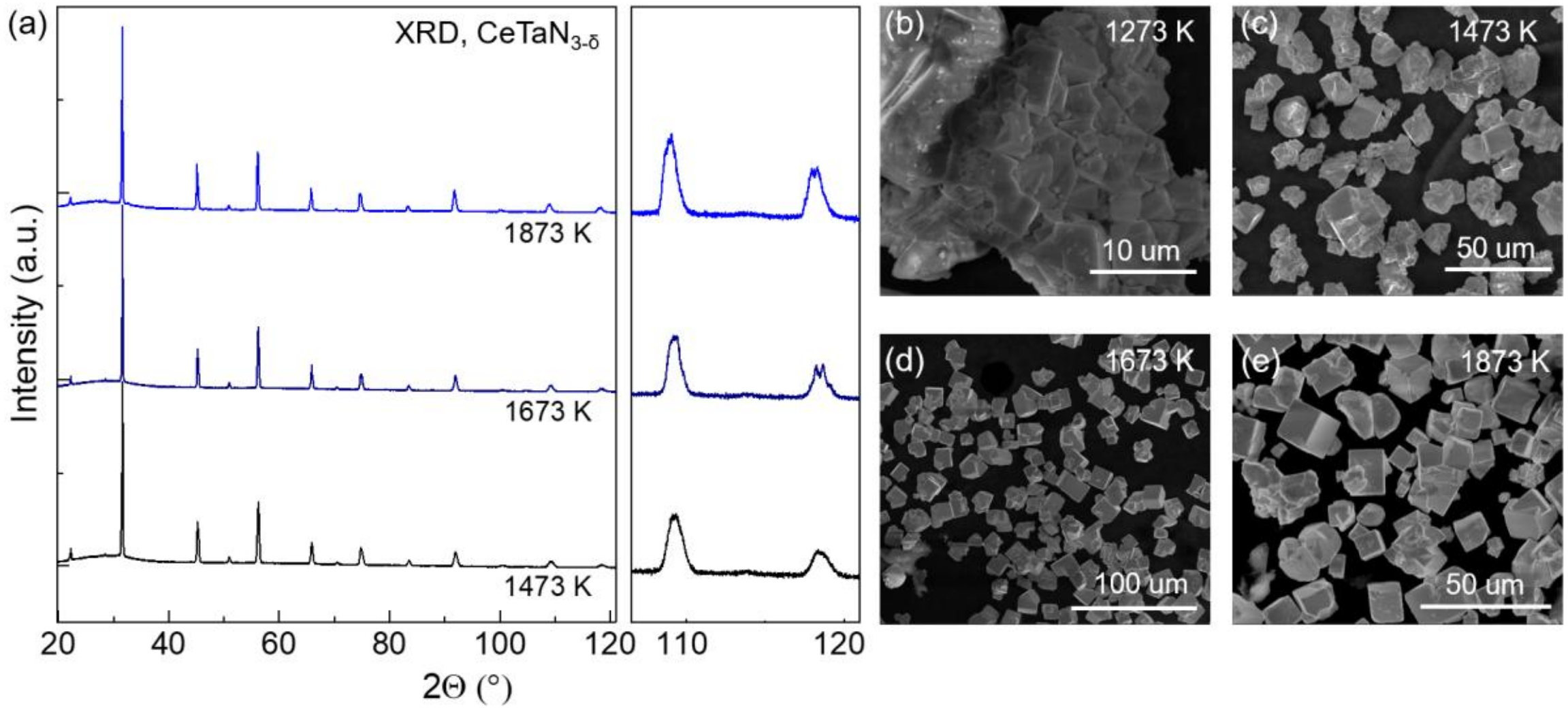

**Fig. S1. XRD patterns and SEM images of as-synthesized $CeTaN_{3-\delta}$.** (**a**) XRD patterns of samples prepared at 5 GPa and various temperatures of 1473, 1673, and 1873 K, respectively. By treatment of starting reactants at 5 GPa and various temperatures above 1473 K, the final products display a same set of XRD peaks, which are initially indexed by a cubic perovskite (i.e., $Pm\bar{3}m$) with a lattice parameter of $a$ = 4.0053(9) Å. (**b**)-(**e**) SEM images of samples synthesized at 5 GPa and the associated different temperatures. Using the similar high-P reaction routes, we have actually also obtained oxygen-free $LaWN_{3-\delta}$ [1], further confirming the effectiveness of the involved high-P methods for forming high-quality nitride perovskites. For the sample synthesized at a higher temperature of 1673 K, a clear peak splitting can be seen, especially for the high-2θ region of reflections, due to the improved sample crystallinity. This means that the nitride has a relatively low symmetry, instead of $Pm\bar{3}m$. At 1873 K, high-quality single crystals of $CeTaN_{3-\delta}$ can be grown with a crystallite size of 10-50 μm. The as-grown crystals have a uniform shape of rectangular cuboid. Based on our further experiments, the edges of most the rectangular cuboid crystals are parallel to the [100], [011], and [01$\bar{1}$] crystallographic directions in terms of the $Pmn2_1$ structural model. The facet normal to [100] is the (100) plane and the associated face diagonals are aligned with [010] and [001].

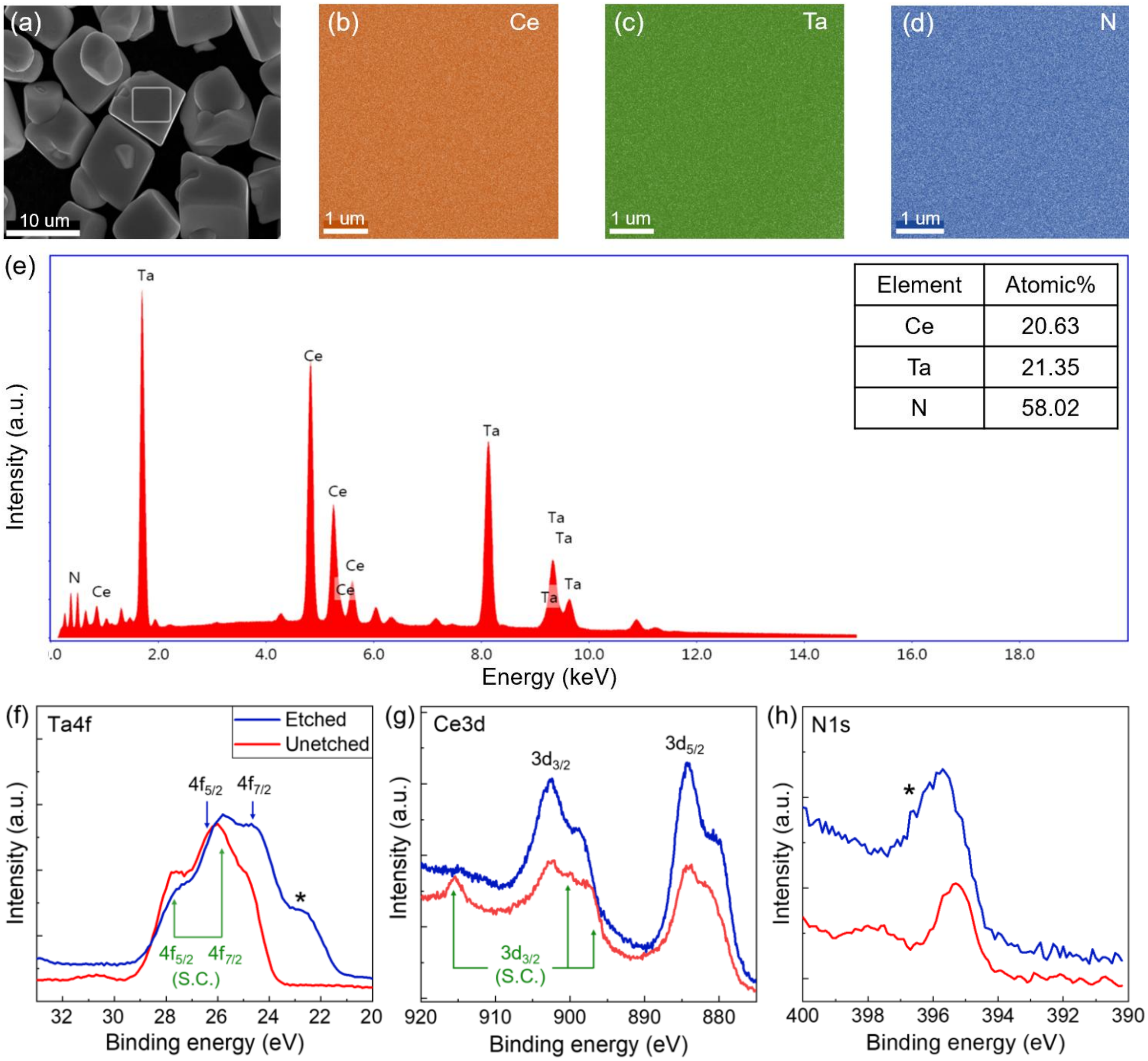

**Fig. S2. EDS and XPS measurements of single-crystal $CeTaN_{3-\delta}$ samples.** (**a**) SEM image. (**b**)-(**d**) Elemental mappings of Ce, Ta, and N, respectively. (**e**) EDS spectrum of a selected area of the sample in (**a**) as framed by the white square line. (**f**)-(**h**) Ta4f, Ce3d, and N1s XPS spectra of unetched and etched crystals, respectively. Green arrows in (f)-(g) denote the spectrum lines with respect to surface contaminations that are nearly inevitable for XPS experiments of many metastable metal nitrides and mostly linked to the surface oxidations of samples in air [1]. The black stars in (**f**) and (**h**) denote the spectrum lines of an unidentifiable Ta-N new phase formed during the etching process, using a high-energy $Ar^{+}$ ion beam with an etching duration of 5 min. Due to the spin-orbit coupling, the associated XPS spectra are split into two main components of $4f_{5/2}$ and $4f_{7/2}$ for Ta and $3d_{3/2}$ and $3d_{5/2}$ for Ce [(**f**)-(**g**)], respectively. For the Ce3d orbitals, both the $3d_{3/2}$ or $3d_{5/2}$ main component lines are further split into doublet peaks (i.e., 898.33 and 902.47 eV for $3d_{3/2}$ and 880.09 and 884.06 eV for $3d_{5/2}$, respectively), which are arguably attributed to the charge-transfer effect [11]. The thus-obtained Ce3d binding energies of $CeTaN_{3-\delta}$ are close to those of $CeF_2$ [12], indicating that the Ce-N bonding state is nearly pure ionic with a $Ce^{3+}$ valence state, rather than the ideal state of $Ce^{4+}$. The determined binding energies of Ta4f are 26.40 eV for $4f_{5/2}$ and 24.50 eV for $4f_{7/2}$, respectively, which are nearly equal to those of $Ta_3N_5$ with excellent mechanical properties [13], implying that the Ta valence state in $CeTaN_{3-\delta}$ is nominally $Ta^{5+}$ and the Ta-N bonds are strongly covalent. Such strong W-N bonding state is an indispensable ingredient for forming robust ferroelectricity with various excellent properties. For the N1s spectrum [(**h**)], an additional peak appears in the etched sample and is located at the higher-energy side, which should originate from the above-mentioned unidentified Ta-N new phase [(**f**)]. As seen in (**f**)-(**g**), for the $Ta4f_{5/2}$ line located at a higher binding energy, there is a profound intensity reduction in the etched sample, indicating that it originates from the surface contaminations. This strongly evidences that the sample is intrinsically oxygen free, although a very thin surface contamination layer may exist and is mostly related to the surface oxidation for many metastable metal nitrides, as also occurred in $LaWN_{3-\delta}$ [1]. Based on both the EDS and XPS measurements, our careful analysis shows that the material is intrinsically oxygen-free and compositionally constituted by Ce, Ta, and N in a molar ratio of ~1:1:3, giving a tentative composition of $CeTaN_3$. Because of the different sensitivities of metals and N to electrons, the EDS method cannot reach an accurate determination of the elemental ratio of most transition-metal nitrides.

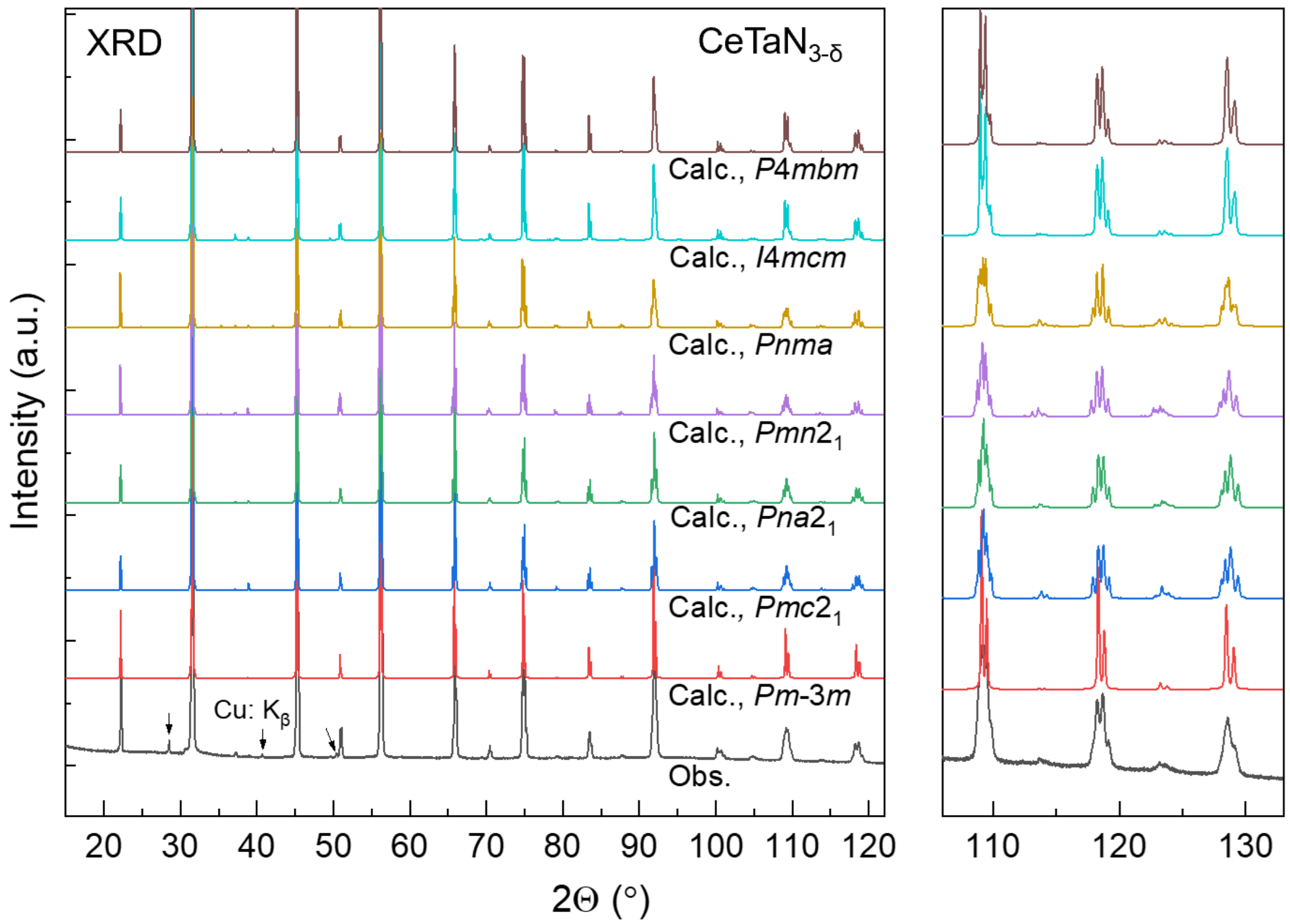

**Fig. S3. Comparison of the observed and calculated XRD patterns for $CeTaN_{3-\delta}$ with various structural models.** The right panel is an enlarged portion to show details. The calculations were performed using various previously proposed structural models of $Pm\bar{3}m$, $Pmc2_1$, $Pna2_1$, $Pmn2_1$, *Pnma*, *I*4*mcm*, and *P*4*mbm* [6]. Before each simulation, the associated structural model is carefully refined based on our XRD and NPD data to obtain more accurate structural parameters for a better simulation. The refined structural details for these structural models are summarized in Table S1. Nearly all the observed XRD reflections lines can be reproduced by each of these structural models.

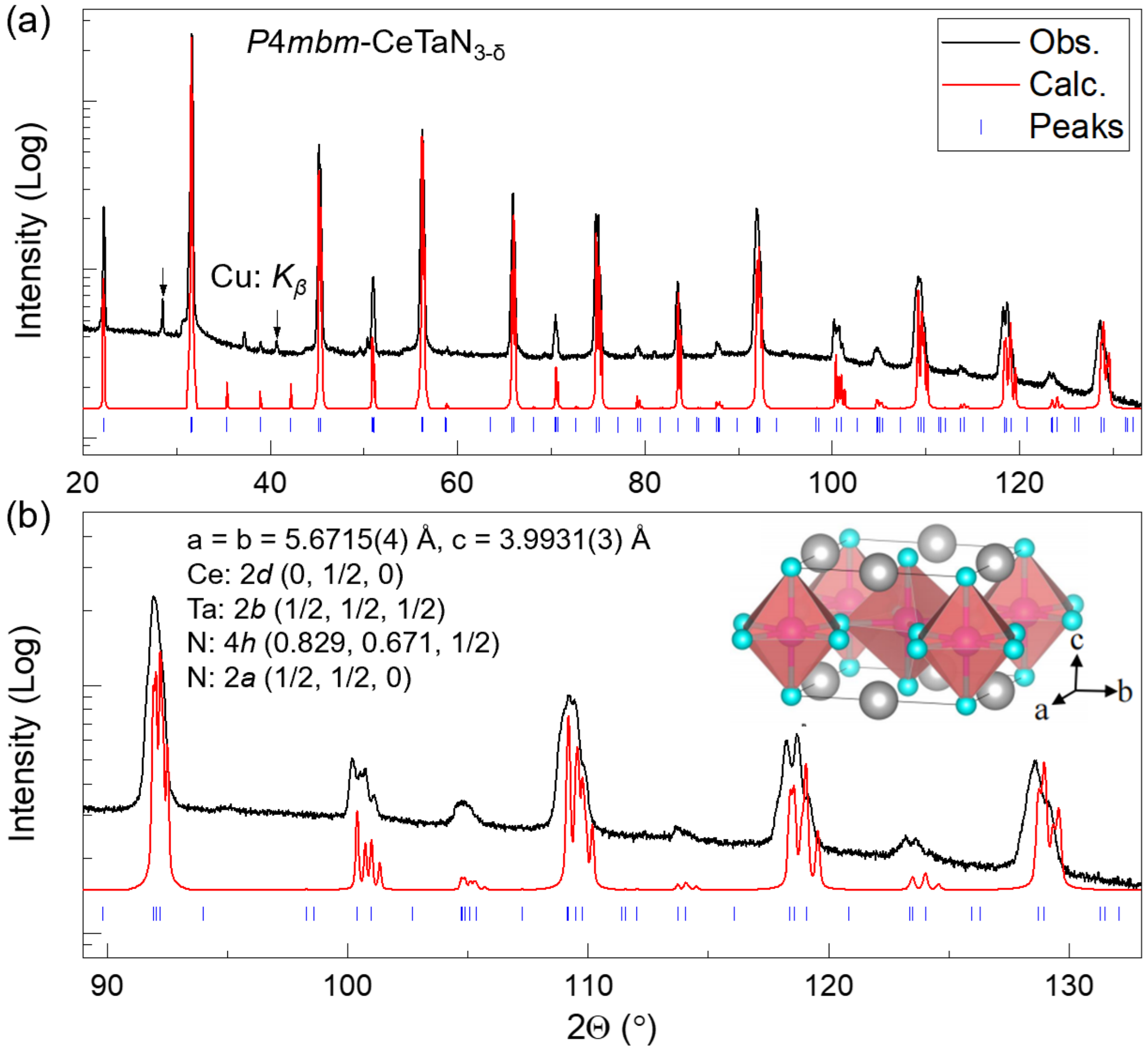

**Fig. S4. Refined XRD pattern using the tetragonal *P4mbm* structural model.** (**a**) Comparison of the observed and simulated XRD patterns. (**b**) Enlarged portion of a high-2θ region pattern. The XRD pattern simulations are performed using a Cu target with involving both the $K\alpha_1$and $K\alpha_2$ radiations (i.e., $\lambda_{K\alpha_1}$= 1.54056 Å and $\lambda_{K\alpha_2}$= 1.54440 Å).

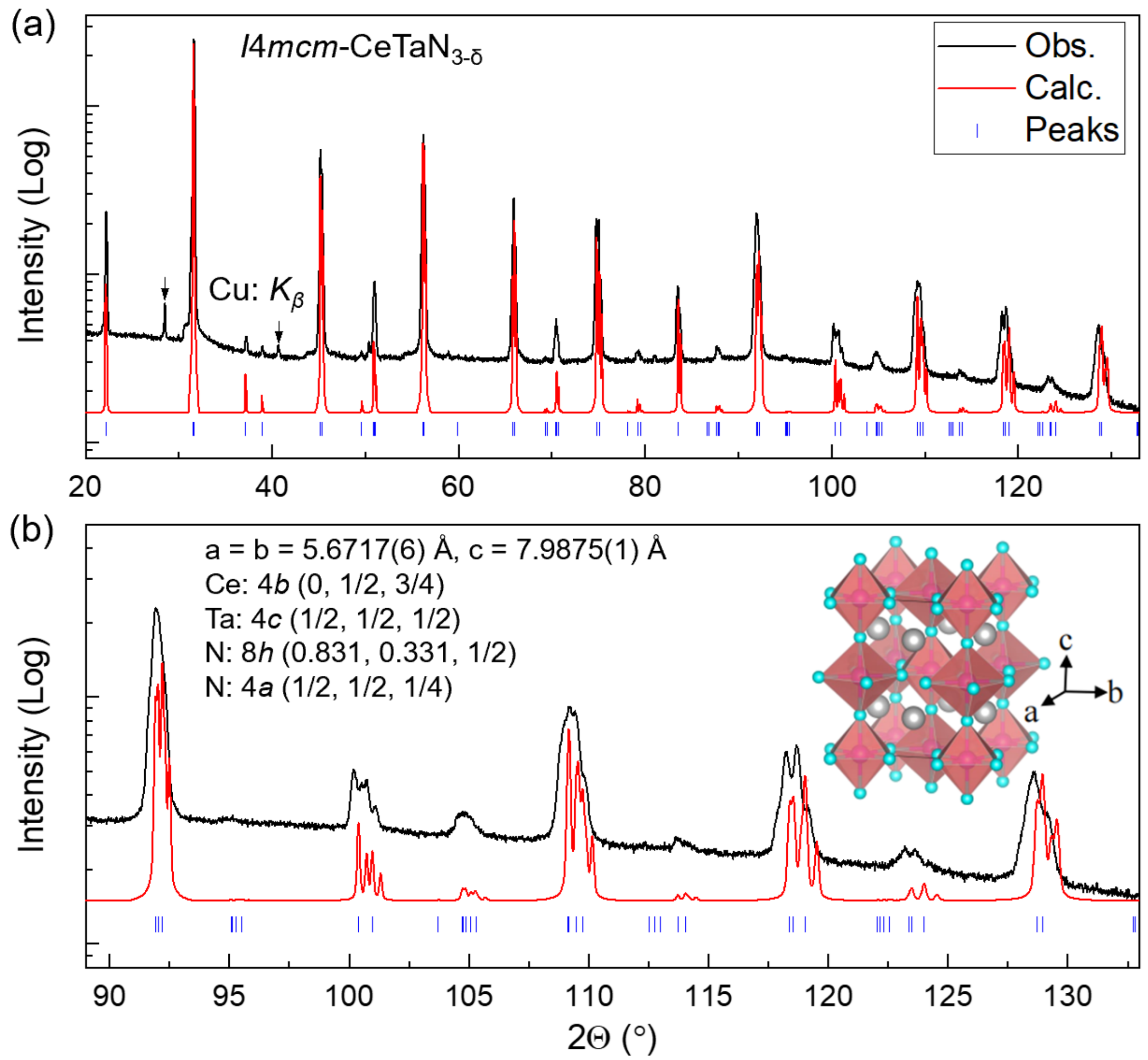

**Fig. S5. Refined XRD pattern using the tetragonal *I4mcm* structural model. (a)** Comparison between the observed and simulated XRD patterns. **(b)** Enlarged portion of a high-2Θ region of the pattern to show details. Inset is a polyhedral view of the refined crystal structure of *I*4*mcm*. The XRD pattern simulations are performed using a Cu target radiation. Upon a closer look of the high-2θ peaks of such simulated XRD patterns, the cases of the two tetragonal structures show obvious discrepancies from the experiments (Figs. S4-S5), by which they can be quickly excluded. For the remaining orthorhombic candidates, a similar excellence of refinement is achieved, implying that they possess an almost same cation sublattice and cannot be differentiated by the x-ray-only data, as the overall XRD signals are predominated by metal atoms, rather than nitrogen.

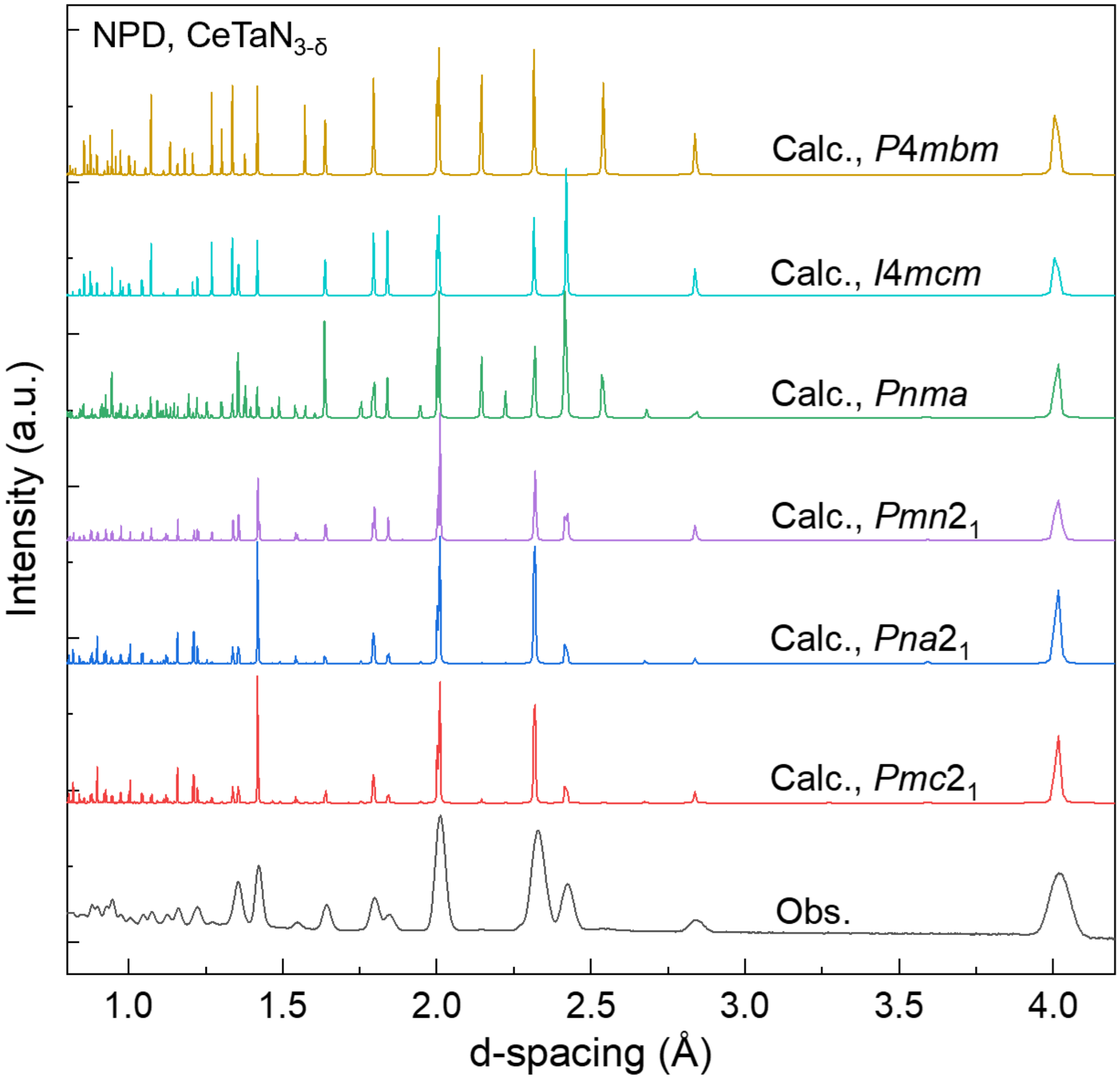

**Fig. S6. Comparison of the observed and simulated NPD patterns.** The simulations are performed using the refined structural models of *Pmc*2$_1$, *Pna*2$_1$, *Pmn*2$_1$, *Pnma*, *I*4*mcm*, and *P*4*mbm*, respectively (see Table S1 for detailed lattice parameters). Similar to the XRD comparisons, the neutron diffraction is a powerful tool to probe N atoms, because the coherent neutron scattering length of N is as large as 9.36 fm, which is almost twice that of Ce (i.e., 4.84 fm) and more than 35% greater than that of Ta (i.e., 6.92 fm) [14]. Comparison of the simulated and observed NPD patterns in Fig. S6 shows that the two tetragonal models present either extra peaks or intensity mismatches and can be readily ruled out.

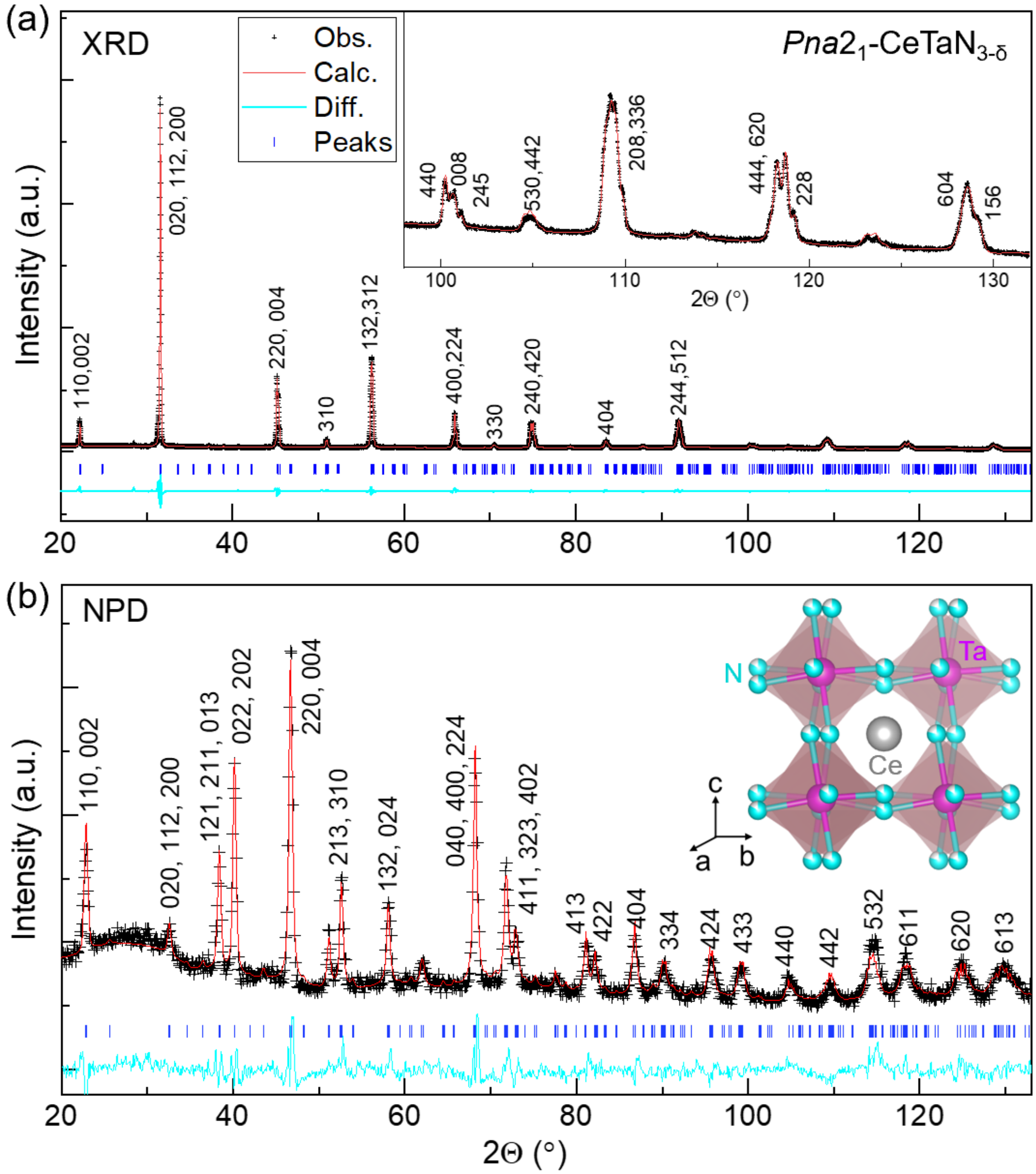

**Fig. S7. Structural refinements of $CeTaN_{3-\delta}$ using the *Pna*$2_1$ structural model.** (**a**) Refined XRD pattern. (**b**) Refined NPD pattern. Inset is a polyhedral view of crystal structure for *Pna*$2_1$-$CeTaN_{3-\delta}$.

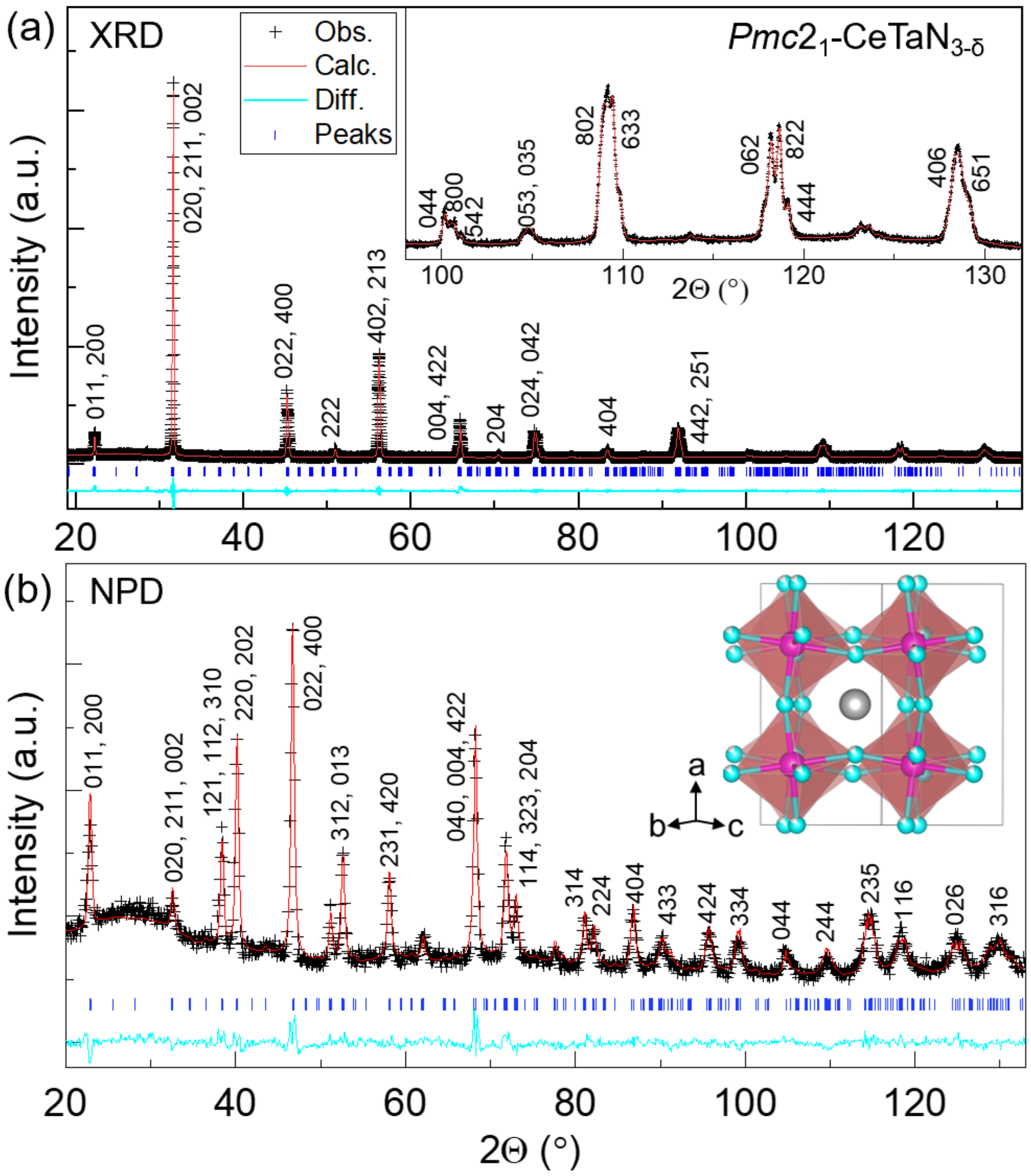

**Fig. S8. Structural refinements of CeTaN$_{3-\delta}$ using the *Pmc*2$_1$ structural model.** (**a**) Refined XRD pattern. (**b**) Refined NPD pattern. Inset is a polyhedral view of crystal structure for *Pmc*2$_1$-CeTaN$_{3-\delta}$

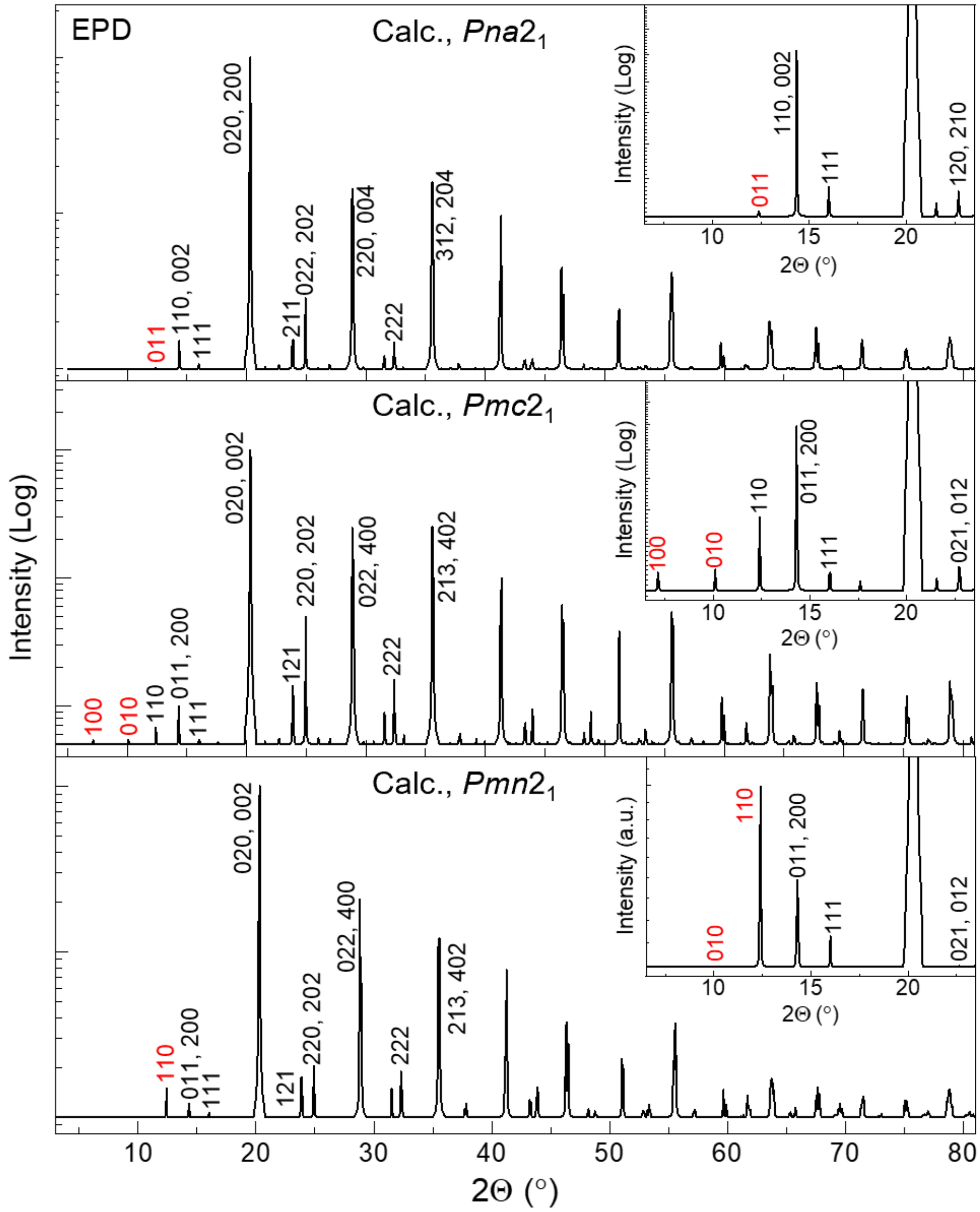

**Fig. S9. Simulated electron powder diffraction (EPD) of $CeTaN_{3-\delta}$ using the refined structural models of *Pmc*2₁, *Pna*2₁, and *Pmn*2₁.** Compared with *Pmc*$2_1$ and *Pna*$2_1$, the EPD simulation of *Pmn*$2_1$ model shows a number of strong diffraction lines (e.g., 101) and very weak lines (e.g., 021 and 012) with nearly negligible intensities. For the case of *Pmc*$2_1$, extra diffraction lines (e.g., 100 and 010) appear. Based on these differences, the crystal structure of $CeTaN_{3-\delta}$ may be distinguished by careful analysis of electron diffraction data.

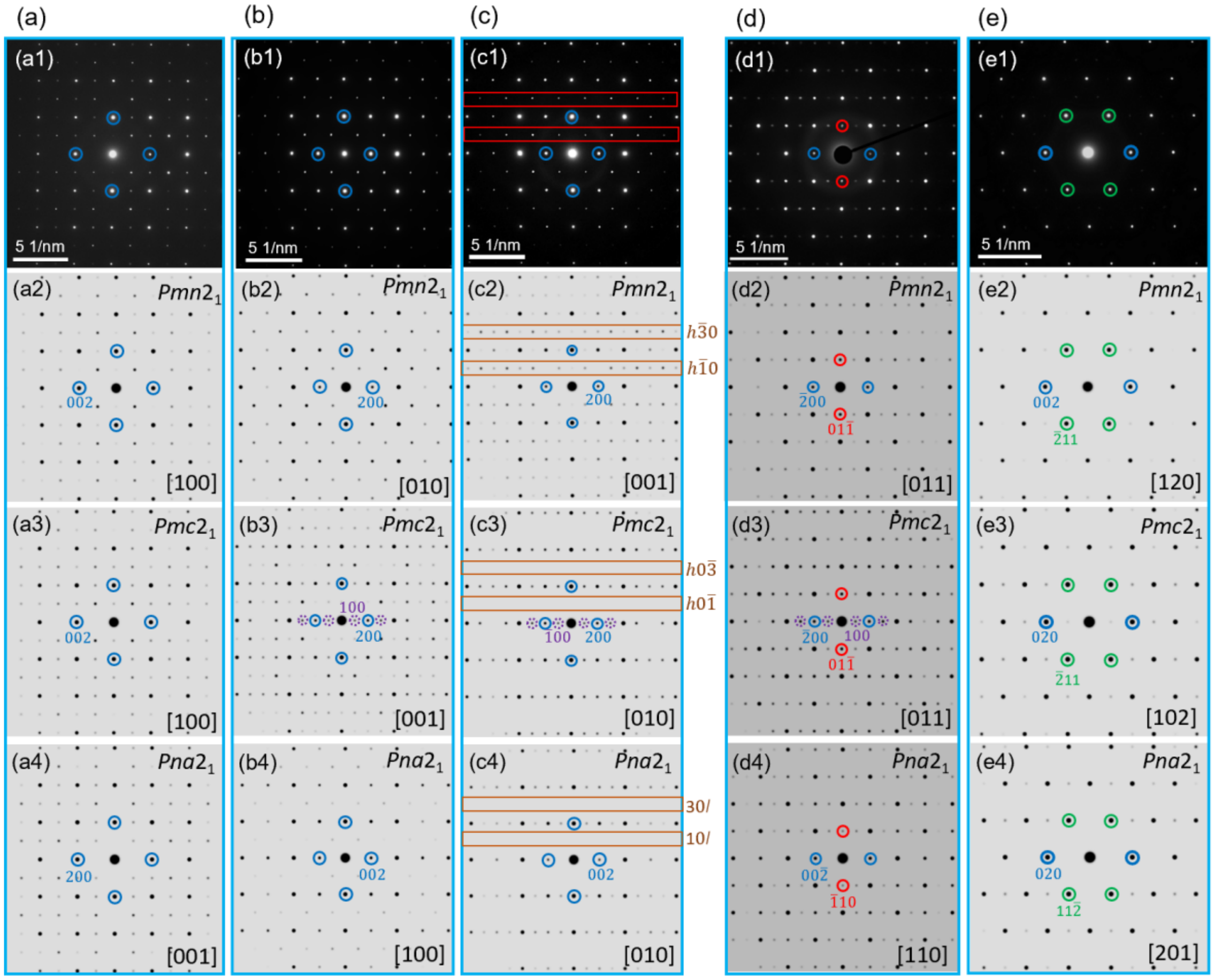

**Fig. S10. Electron diffraction analysis for CeTaN$_{3-\delta}$ to exclude both the competing *Pmc*2$_1$ and *Pna*2$_1$ models. (a)-(e) Observed and simulated** SAED patterns of CeTaN$_{3-\delta}$ along five different zone axes of [100], [010], [001], [011], and [120] as assigned in terms of the *Pmn*2$_1$ model. All the SAED patterns are taken with an aperture of 400 nm in size. **(a1) - (e1)** Observed SAED patterns. **(a2) - (e2)** Simulated SAED patterns using the refined *Pmn*2$_1$-CeTaN$_{3-\delta}$. **(a3) - (e3)** Simulated SAED patterns using the refined *Pmc*2$_1$-CeTaN$_{3-\delta}$. **(a4) - (e4)** Simulated SAED patterns using the refined *Pna*2$_1$-CeTaN$_{3-\delta}$. Along the $[100]_{Pmn2_1}$ and $[120]_{Pmn2_1}$ directions, the three models of *Pmn*2$_1$, *Pmc*2$_1$, and *Pna*2$_1$ give nearly distinguishable SAED simulations, consistent with the observed in **(a)** and **(e)**. The models of *Pmn*2$_1$ and *Pna*2$_1$ cannot be differentiated from each other along the zone axes of $[010]_{Pmn2_1}$ and $[011]_{Pmn2_1}$, as seen in **(b)** and **(d)**. However, along the $[010]_{Pmn2_1}$, $[001]_{Pmn2_1}$, and $[011]_{Pmn2_1}$ directions, the *Pmc*2$_1$ model produces extra diffraction spots [such as 100 in **(b3) - (d3)**] and systematic extinction of certain rows of diffraction spots [e.g., $h0\bar{1}$ and $h0\bar{3}$) in **(c3)**]. Using the *Pna*2$_1$ model, the similar extinction of diffraction-spot rows can also be seen in **(c4)**. These features cannot be observed in the experiments in **[(b1) - (d1)]**, by which both *Pmc*2$_1$ nor *Pna*2$_1$ can be easily ruled out. By contrast, all the experimental SAED patterns can be excellently reproduced by the *Pmn*2$_1$ model [**(a2) - (e2)**], strongly implying that the suitability of *Pmn*2$_1$ for well describing the crystal structure of CeTaN$_{3-\delta}$. We thus conclude that *Pmn*2$_1$ is the most appropriate crystal structure for this nitride and used for the subsequent structural resolution.

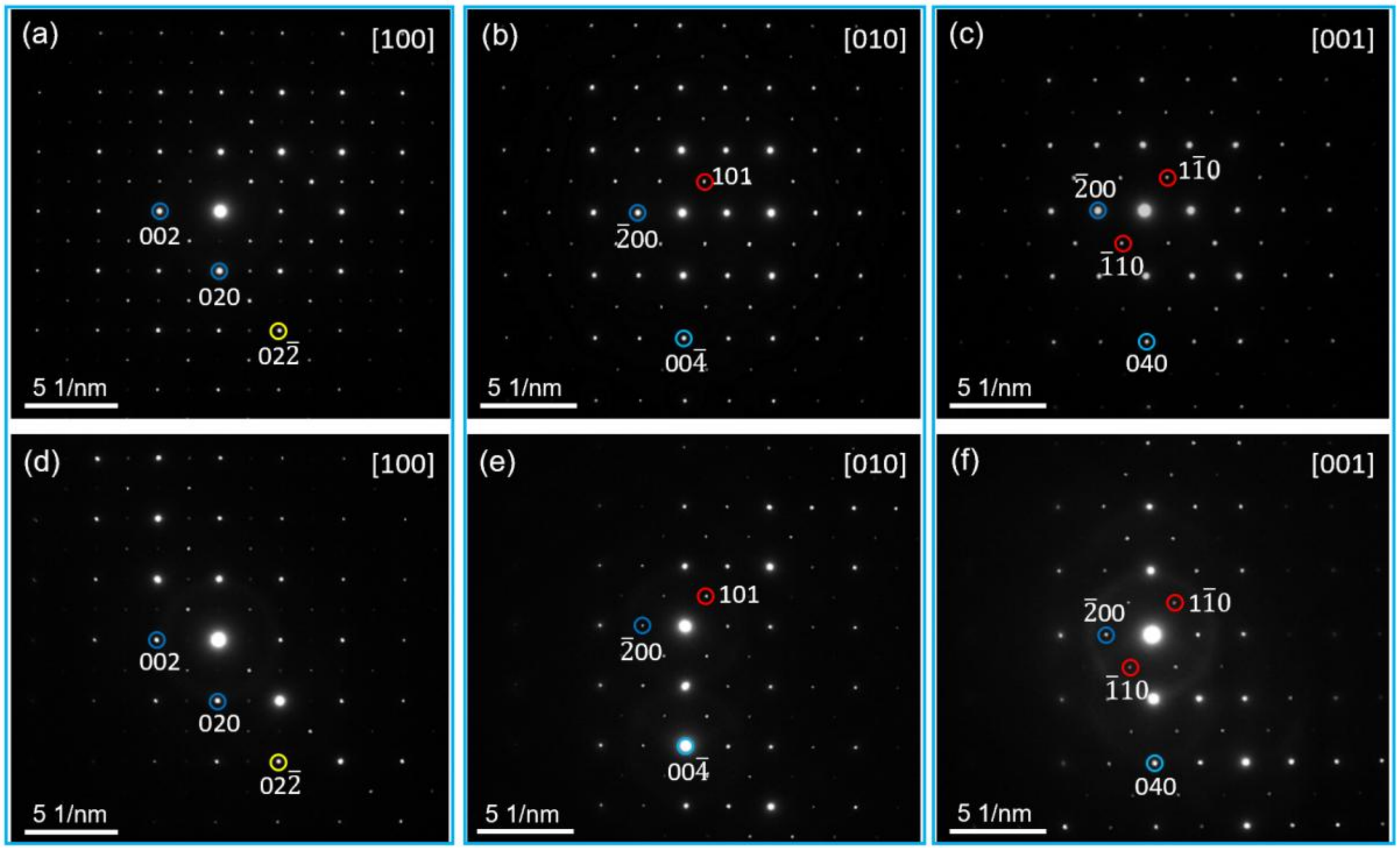

**Fig. S11. SAED patterns of *Pmn*$2_1$-$CeTaN_{3-\delta}$ collected along both the untilted and tilted zone axes of [100], [010], and [001]. (a)-(c)** Along the untilted zone axes. **(d)-(f)** Along the tilted zone axes. Apparently, by tilting the zone axes with a tilting angle of ~2°, the thus-obtained SAED patterns show that all the diffraction spots are retained, indicating they arise from the primary diffraction without involvement of high-order diffraction spots[15].

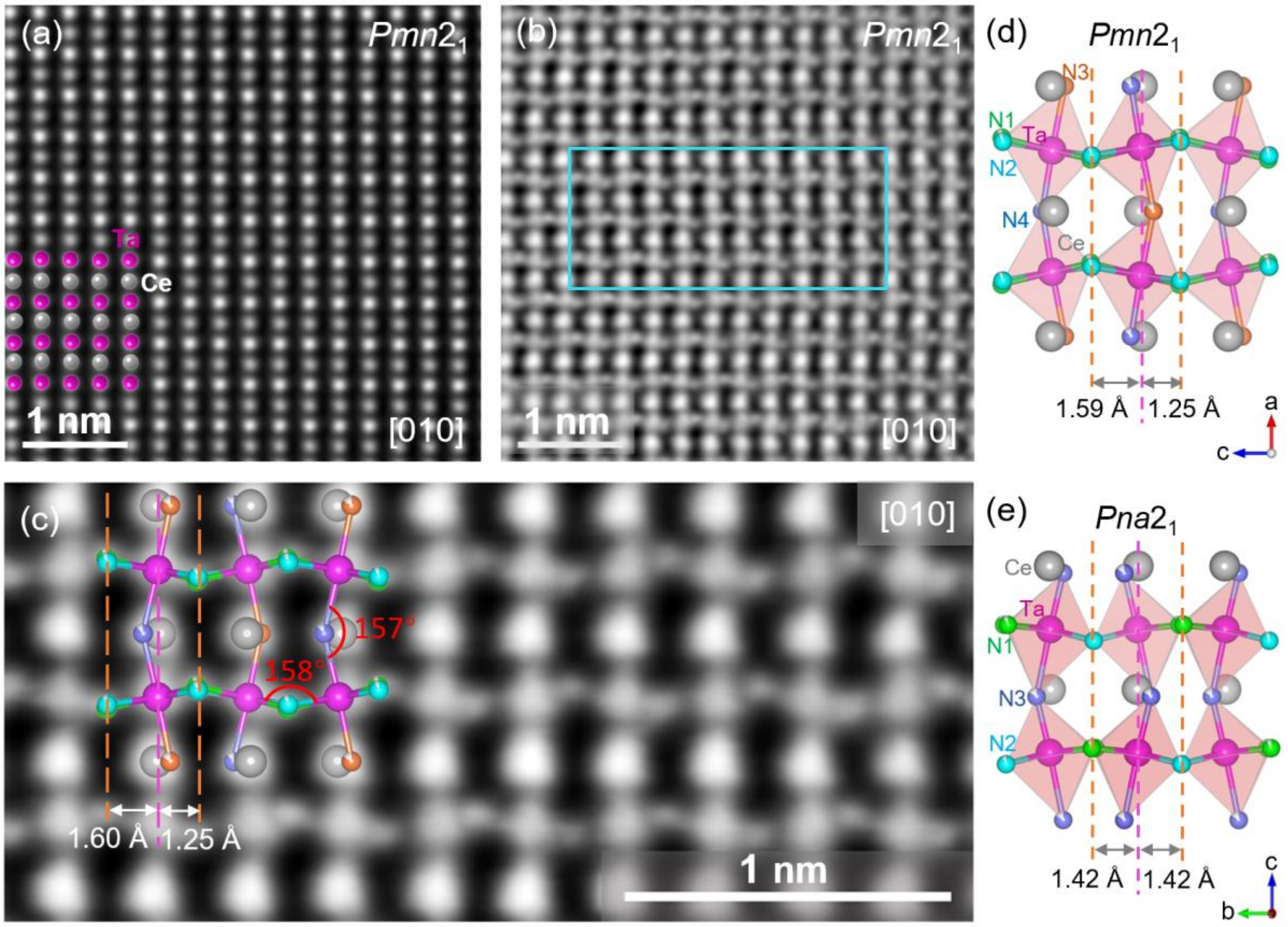

**Fig. S12. Structural identification of $CeTaN_{3-\delta}$ by STEM observations. (a)** HAADF-STEM image. **(b)** iDPC-STEM image. **(c)** Enlarged portion of iDPC-STEM image as framed in (b). Inset is the crystal structure of *Pmn*2$_1$-$CeTaN_{3-\delta}$. **(d)-(e)** Refined crystal structures of *Pmn*2$_1$ and *Pna*2$_1$ viewed along the $[010]_{Pmn2_1}$ and $[100]_{Pna2_1}$ directions, respectively. The model of *Pna*2$_1$ predicts a zero displacement of Ta atom relative to the center of its surrounding nitrogen octahedral coordination along the $[010]_{Pna2_1}$ direction **(e)**, which is inconsistent with the atomic positions determined from iDPC-STEM observations in **(c)**. This, combined with comprehensive SAED observations in Fig. S11, fully exclude the *Pna*2$_1$ model.

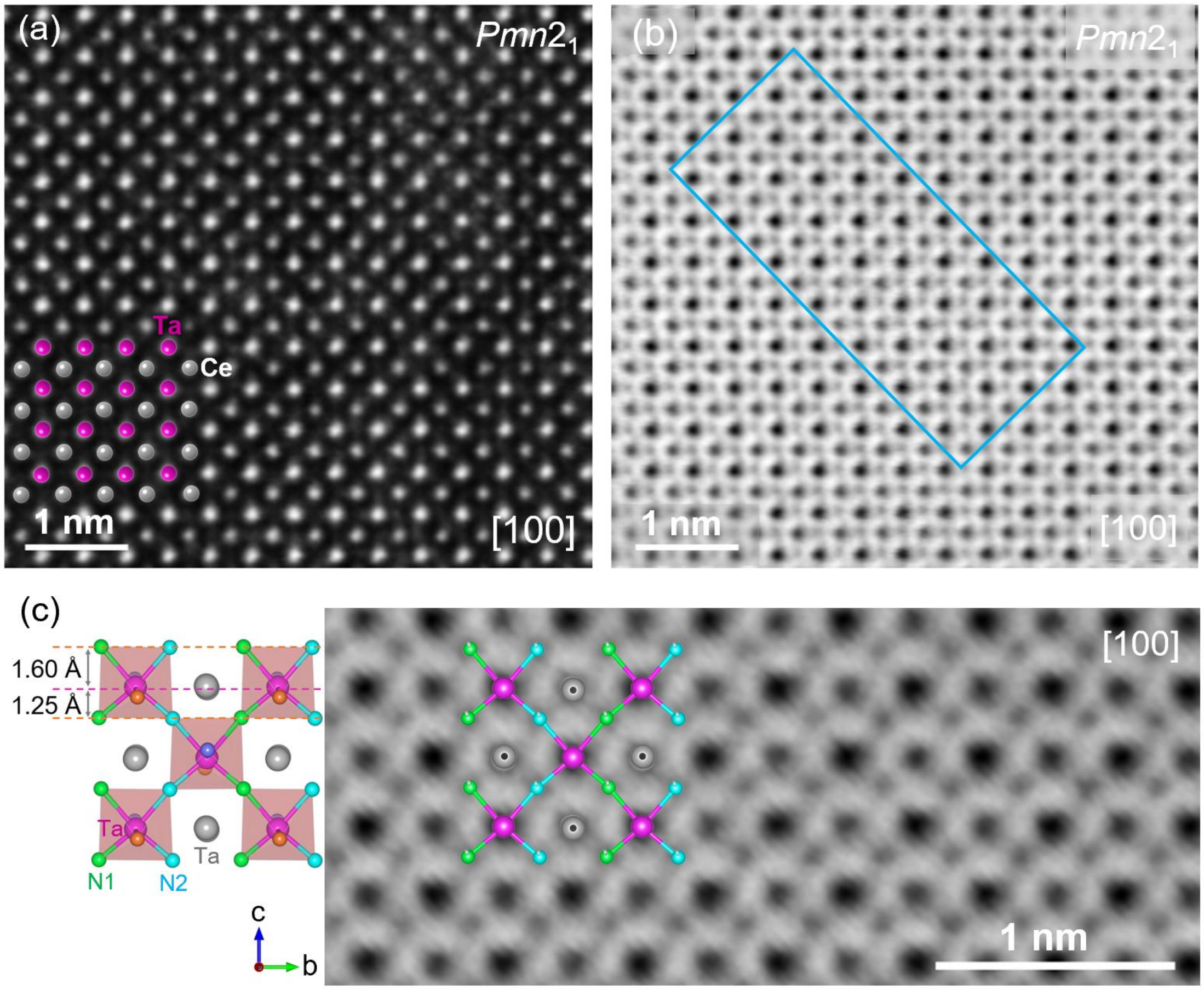

**Fig. S13. Atomic-scale observations of *Pmn*2$_1$-CeTaN$_{3-\delta}$ along the $[100]_{Pmn2_1}$ crystallographic direction by STEM experiments. (a)** HAADF-STEM image. **(b)** ABF-STEM image. **(c)** Enlarged portion of ABF-STEM image as framed in (b). Inset is the crystal structure that is mapped on the atomic lattice fringes. It can be seen that the Ta atom of each {$TaN_6$} octahedron has a clear shift relative to its N coordinates along the [001] direction. The thus-measured relative shift of Ta is $\Delta r \approx 0.185$ Å by ABF-STEM observation, which is very close to the refined value of ~0.175 Å based on the structural analysis (see Fig. 2 of the main text).

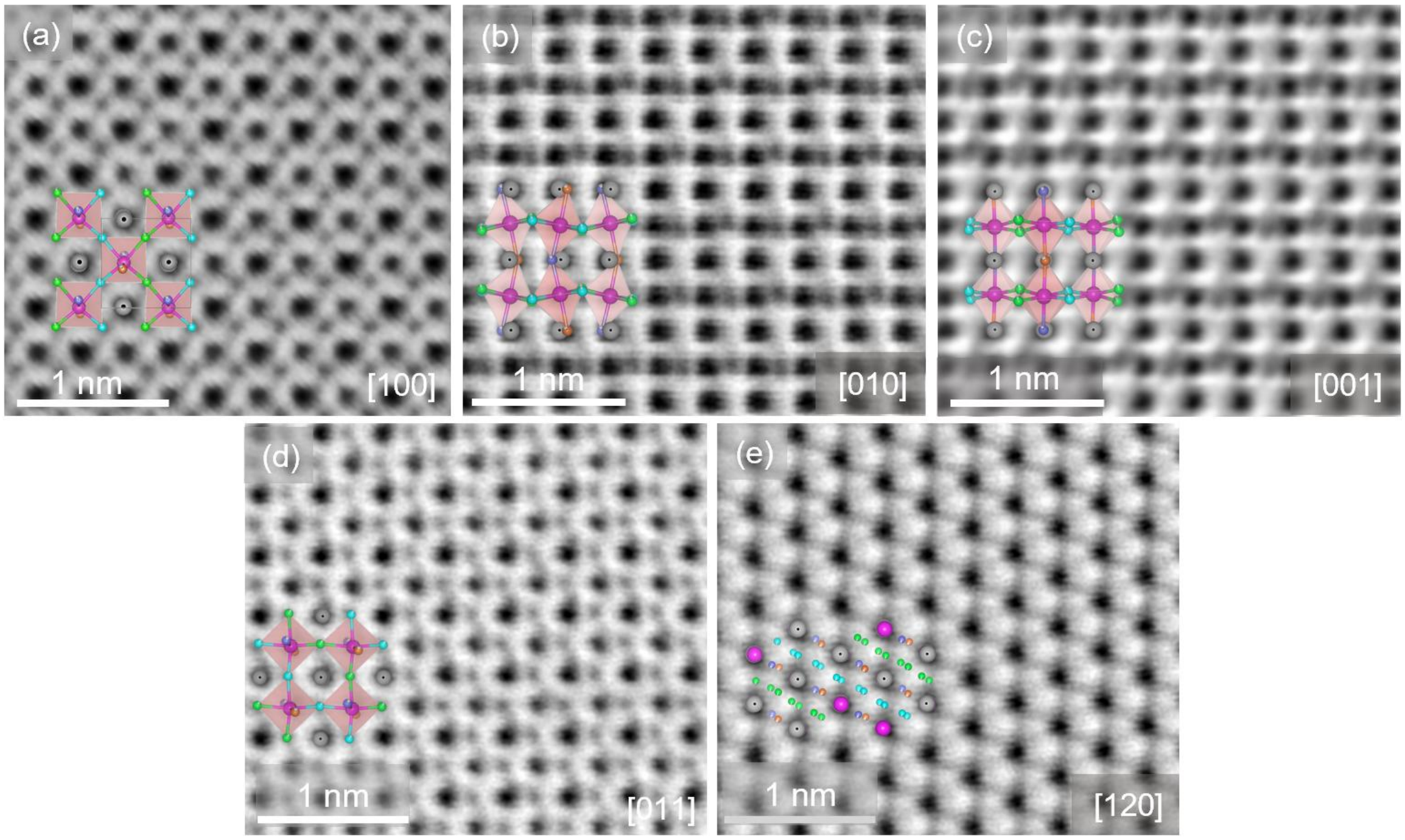

**Fig. S14. ABF-STEM observations of *Pmn*2$_1$-CeTaN$_{3-\delta}$ along five different zone axes of [100], [010], [001], [011], and [120], respectively.** Inset in each panel is the crystal structure of *Pmn*2$_1$-CeTaN$_{3-\delta}$ that is mapped on the ABF image.

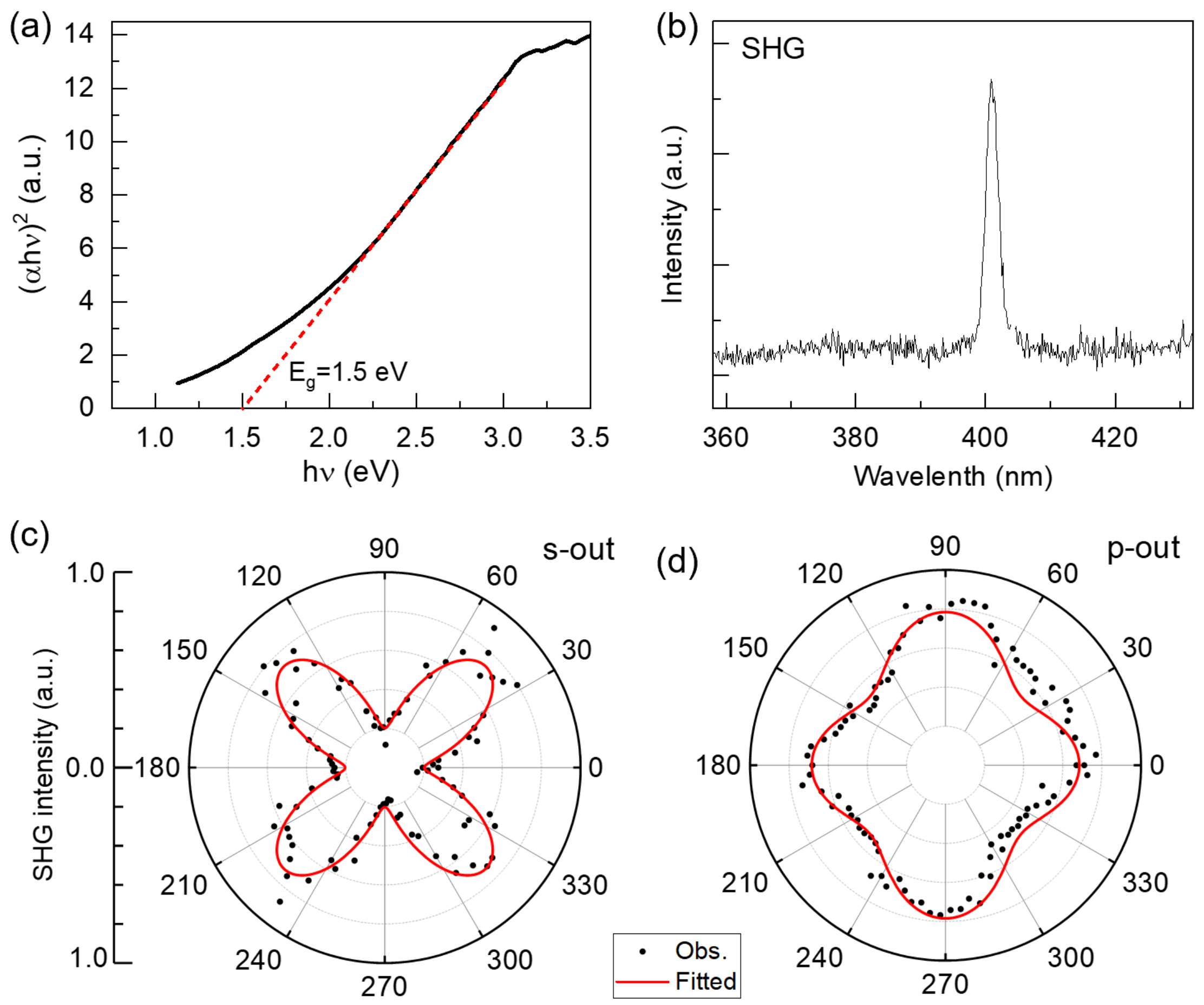

**Fig. S15. Spectroscopic measurements of *Pmn*2$_1$-CeTaN$_{3-\delta}$. (a)** Ultraviolet-visible absorption spectrum. (**b**) Second-harmonic generation (SHG) spectrum. (**c**)-(**d**) Angle-dependent SHG spectra. *p*-out and *s*-out denote the involved two orthogonal directions. The least-squares analysis of the obtained SHG data shows that both the outgoing *s*-out (i.e., $I_\perp$) and *p*-out (i.e., $I_{||}$) signals can be well described by $I_\perp = I_0 sin^2 2\varphi + A$ and $I_{||} = I_0 cos^2 2\varphi + B$, respectively, with fitted constants of $I_0$ = 52.8, $A$ = 20.2, and $B$ = 130.5 in a same unit of mW/mm$^2$. Variables $A$ and $B$ primarily originate from the electric-domain effects, accounting for the observed nonzero background signals along the two orthogonal directions; the discrepancy between them indicates a preferred domain orientation of the crystal.

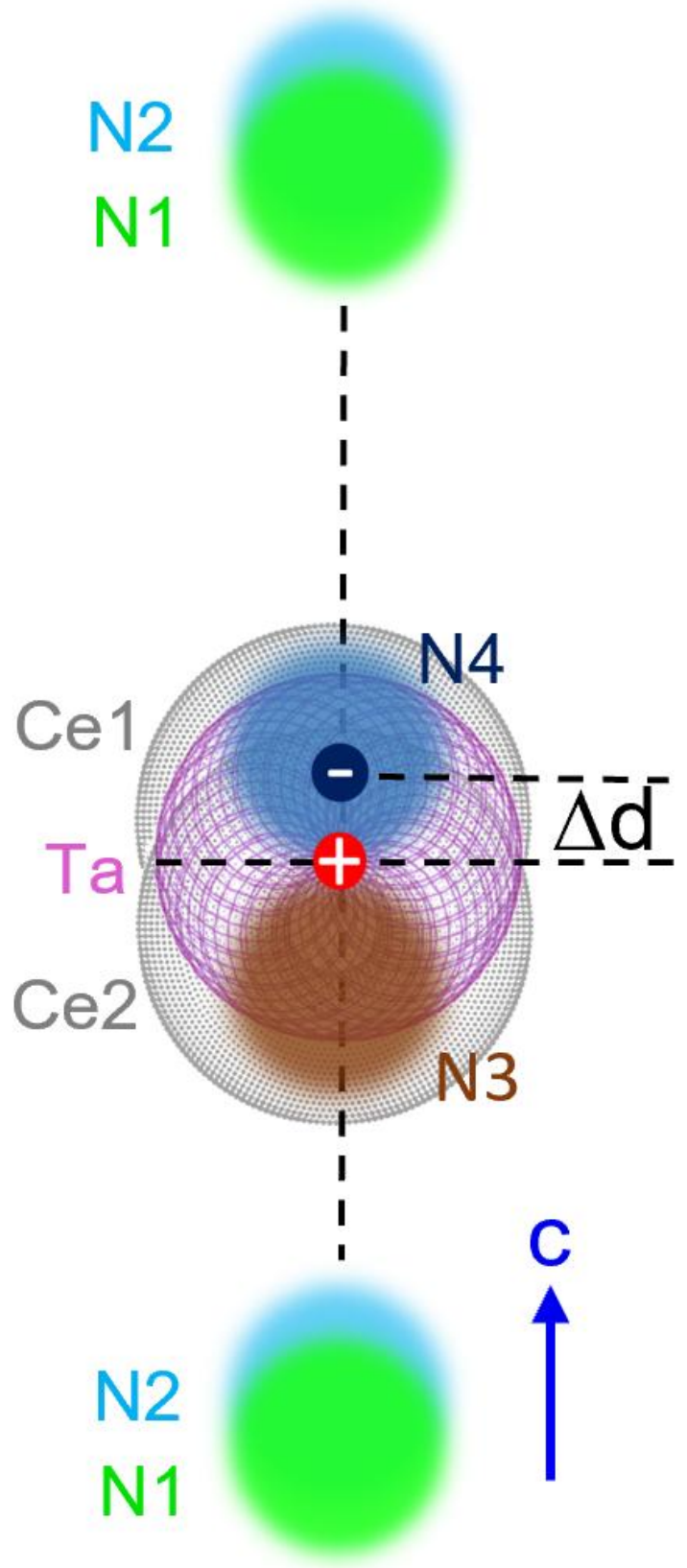

**Fig. S16. Projection of the reduced unit on the *c*-axis.** An effective electric dipole is produced along the *c*-axis by the formation of equivalent negative (i.e., all six N anions) and positive (i.e., one Ta and two Ce cations) charge centers of $Z_-$ and $Z_+$, which can be calculated by

$$Z_{\pm} = \frac{\sum(N_i \eta_i) Z_i}{\sum(N_i \eta_i)} .$$

Where $N_i$ is the ionic valence state (*i.e.,* +3, +5, and -3 for the involved ions of $Ce^{3+}$, $Ta^{5+}$, and $N^{3-}$, respectively), $\eta_i$ is the refined atomic occupancy, and $Z_i$ is the atomic position vector projected along the polarization direction (i.e., the *c*-axis for *Pmn*$2_1$-$CeTaN_{3-\delta}$). The refined $\eta_i$ and $Z_i$ values can be found in Table 1 of the main text. The thus-obtained displacement between negative and positive charge centers are $\Delta d = Z_- - Z_+ \approx 0.1514$ Å at ambient conditions. The theoretical polarization of $P_{theo.}$ can be readily calculated to be ~29.9 μC/cm$^2$ by using Eq. S3 [i.e., $P_{theo.} = \frac{1}{V}(-e\frac{\sum(N_i\eta_i)Z_i}{\sum(N_i\eta_i)})$], which is larger than that of the case without considering the atomic deficiencies with a value of ~28.8 μC/cm$^2$. This indicates that influence of atomic deficiencies on electric polarization is small. Besides, the two Ce ions are symmetrically positioned about the Ta ion and should have a negligible contribution to electric polarization. Indeed, our calculations show that polarization with only the Ta ions yields a value of 29.3 μC/cm$^2$, marginally different from that of the case involving both Ce and Ta ions (i.e., 29.9 μC/cm$^2$).

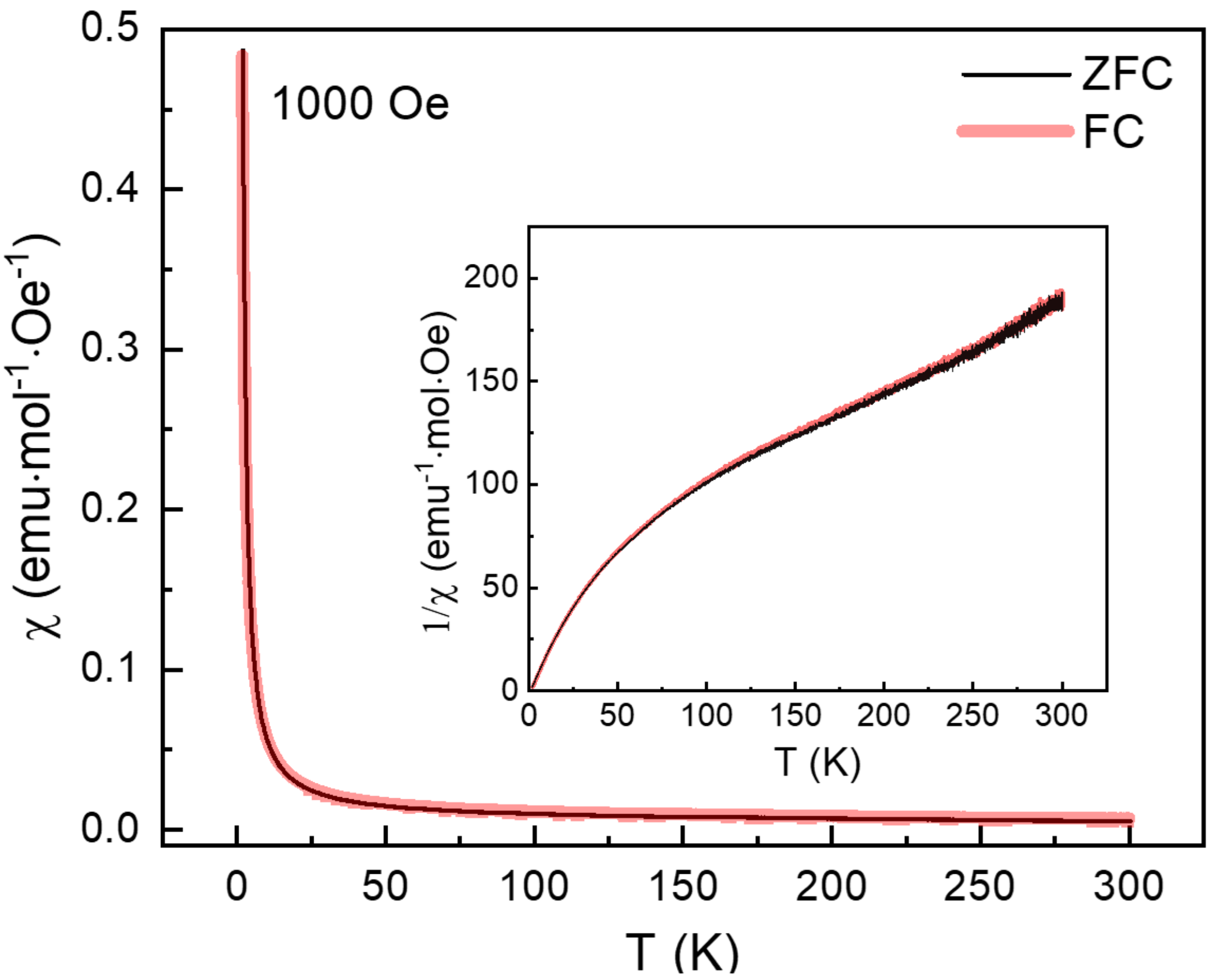

**Fig. S17. Low-T magnetic susceptibility (χ) of $CeTaN_{3-\delta}$ measured at 1000 Oe under zero-field cooled (ZFC) and field cooled (FC) conditions, respectively.** Inset shows the inverse susceptibility ($\chi^{-1}$).

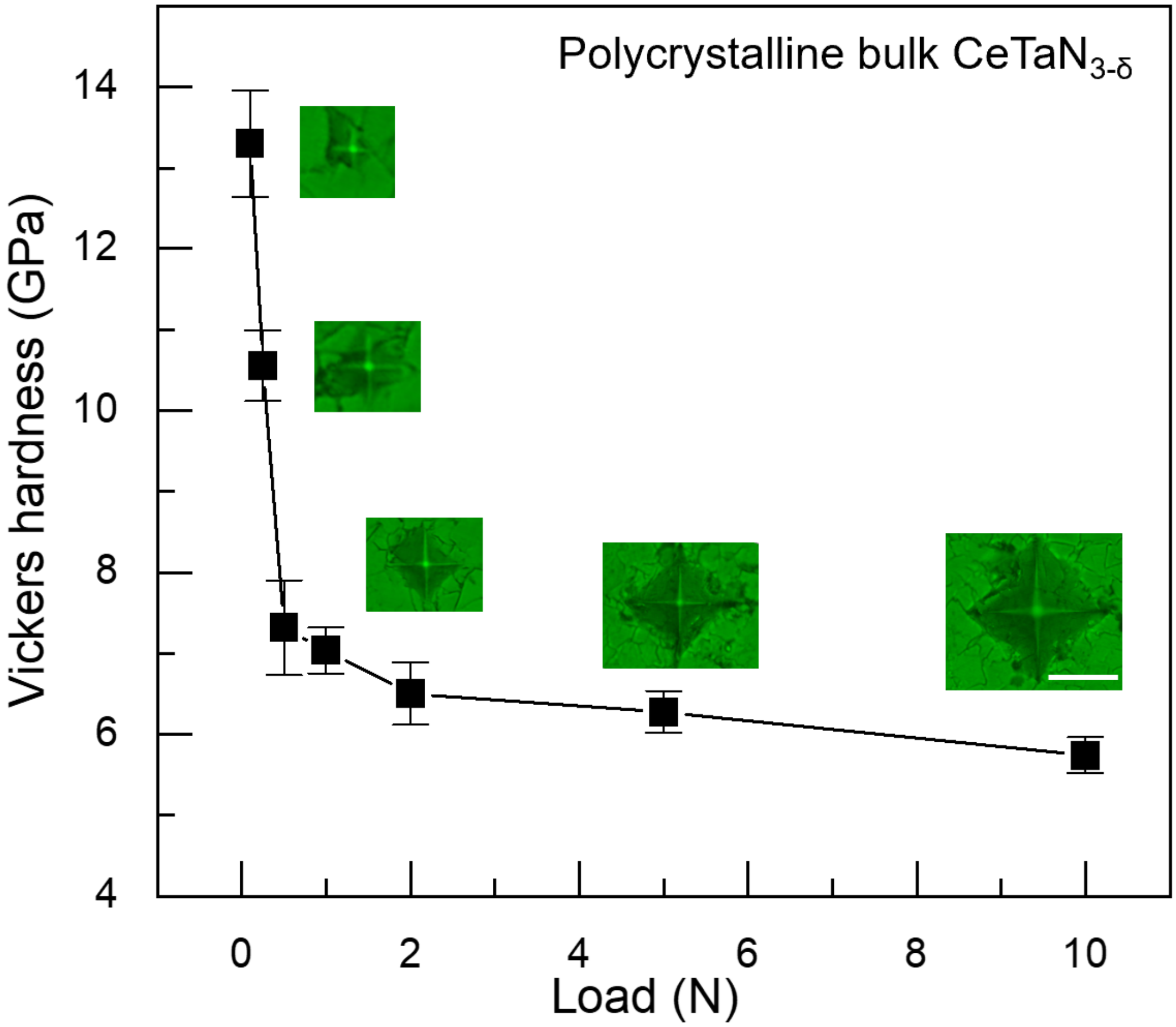

**Fig. S18. Vickers hardness measurements of a well-sintered polycrystalline bulk $CeTaN_{3-\delta}$ sample.** The bulk sample was sintered at 10 GPa and 1473 K for 30 min, using the synthesized powders as starting materials. Insets are typical images of indentations under different loads, and the scale bar is 20 μm. Apparently, the load-invariant hardness of this material is about 6 GPa, comparable to that of $LaWN_{3-\delta}$ [1]. However, the occurrence of many cracks around the indentations indicates a low toughness of this nitride, characteristic of a brittle ceramic.

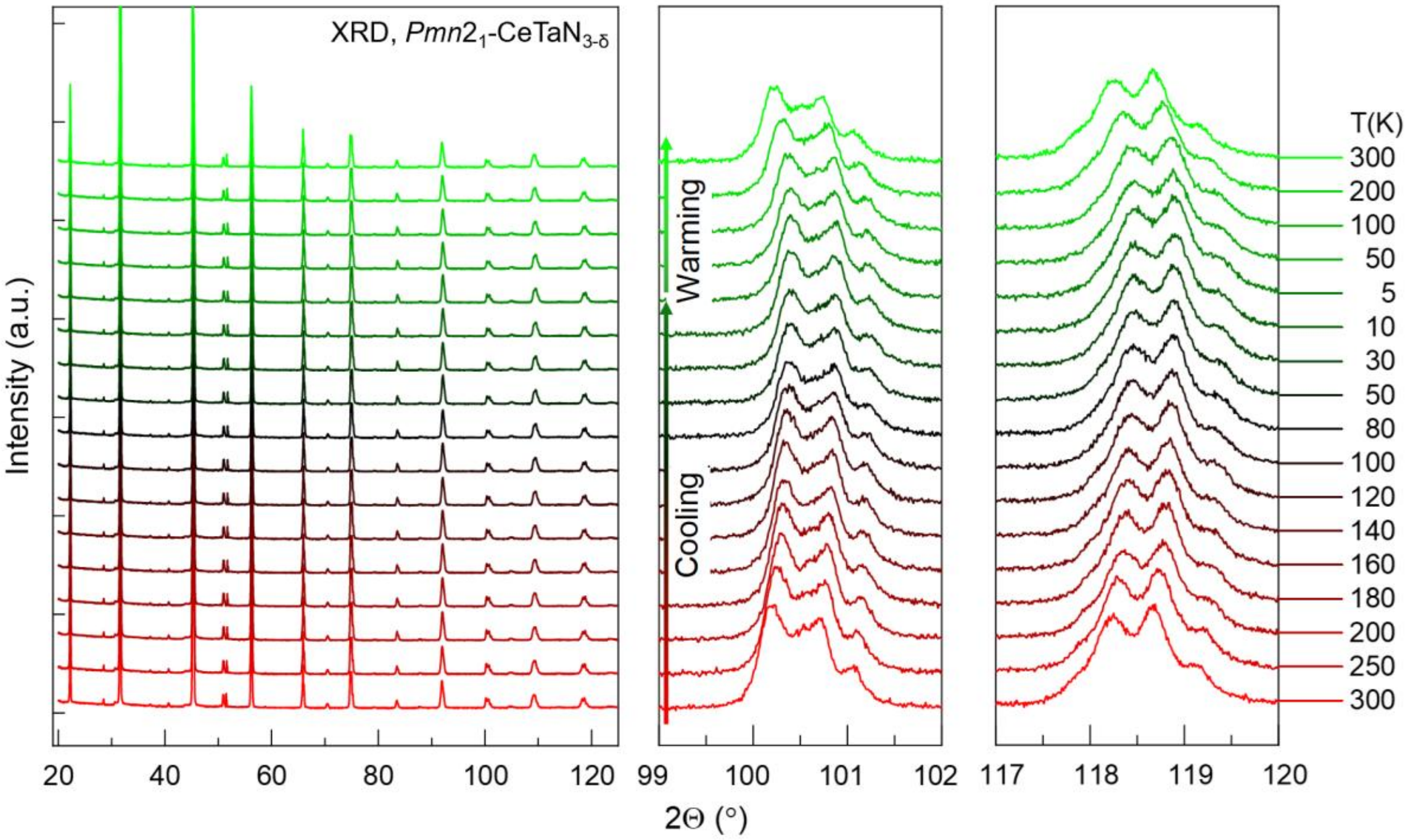

**Fig. S19. Low-T XRD patterns of $CeTaN_{3-\delta}$ taken on cooling.** The right two panels are enlargements to show the details of peak profiles around 2Θ = 101° and 118°, respectively. Low-T XRD experiments are carried out in the 5-300 K temperature range at ambient pressure to explore structural stability at low temperatures. To reduce the temperature fluctuation, at each desired temperature the sample is soaked for 10 min before data collection. The sample position is also programmable realigned to obtain strong diffraction signals.

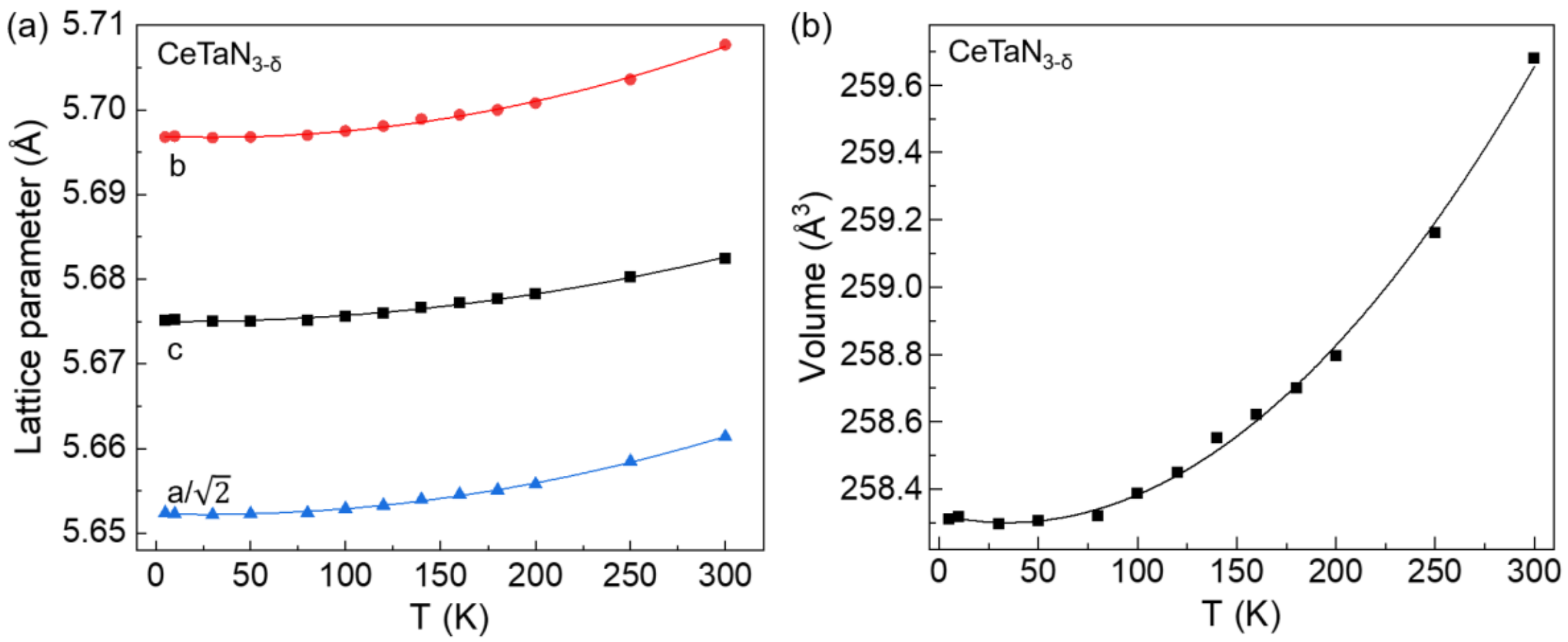

**Fig. S20. Refined lattice parameters vs. temperature for *Pmn*2$_1$-CeTaN$_{3-\delta}$, based on low-T XRD measurements.** (**a**) Lattice parameters of *a*, *b*, and *c* with varying temperature. (**b**) Unit-cell volume vs. temperature.

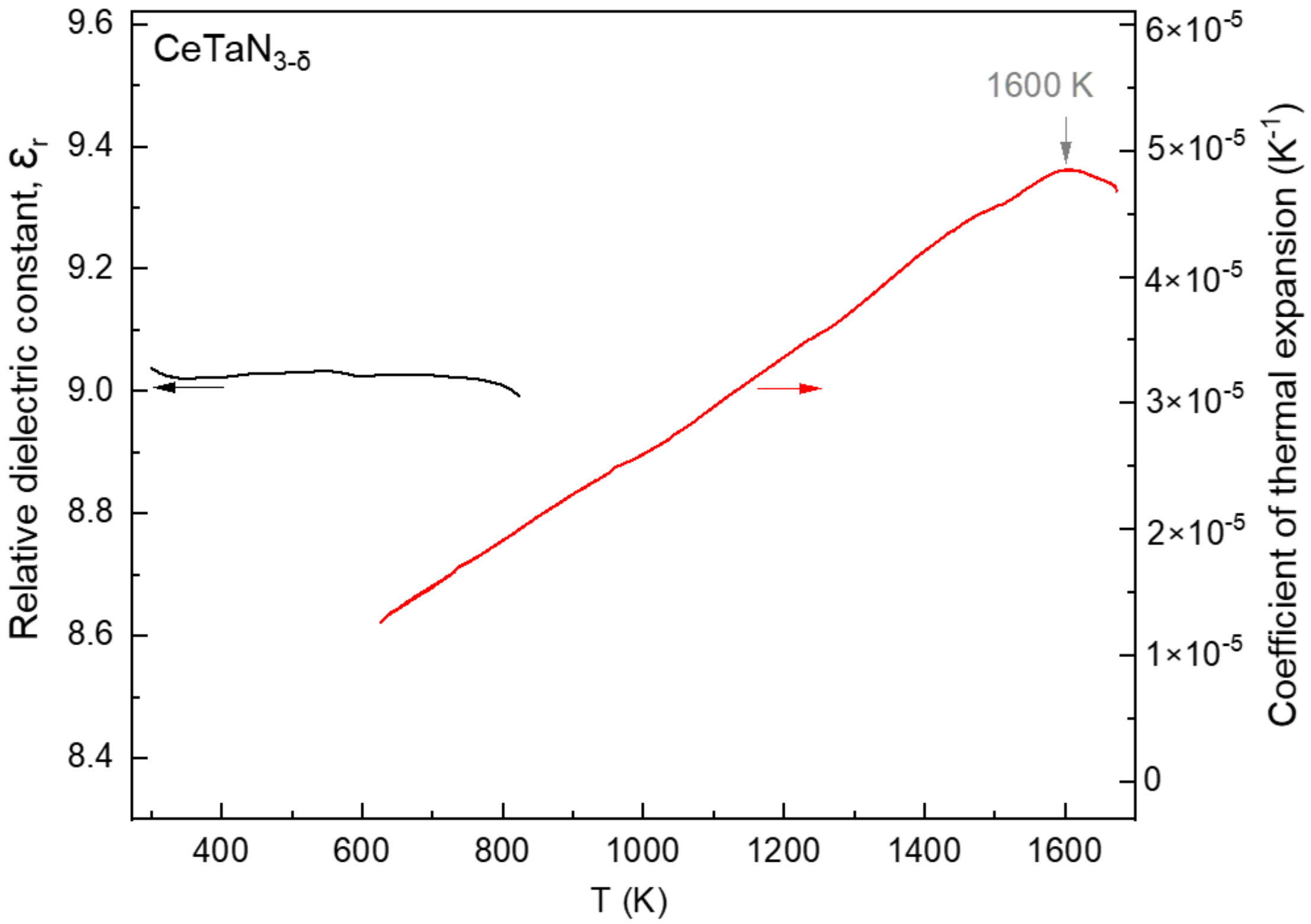

**Fig. S21. Temperature-dependent relative dielectric constant ($\varepsilon_r$) and coefficient of thermal expansion (CTE) of $CeTaN_{3-\delta}$.** The dielectric constant ($\varepsilon$) measurement is performed at 1000 Hz and the highest achievable temperature of the involved instrument is 800 K. The relative dielectric constant is calculated by $\varepsilon_r = \varepsilon/\varepsilon_0$, where $\varepsilon_0 \approx 8.854 \times 10^{-12}$ F/m represents the dielectric constant of vacuum. Because of a large instrumental uncertainty of CTE measurement below 600 K, the experiment is conducted in the 600-1673 K temperature range to obtain reliable experimental data. The arrow denotes the onset decomposition temperature of 1600 K and the final products are $Ta_2N$ and CeN as identified by XRD measurement. Compared with many nitrogen-rich binary transition-metal nitrides with low decomposition temperatures (e.g., 1173 K for $Ta_3N_5$ and 850 K for $MoN_2$) [16,17], the sample of $CeTaN_{3-\delta}$ starts to decompose at a high temperature of 1600 K, indicating its superior stabilities to withstand high temperatures. This is also related to the strong covalent Ta-N bonding, due to the presence of Ce with a low electronegativity that functions as an excellent electron donator for enhancing the Ta-N covalency [18]. Also noted is that the sample's stability can further be promoted above 2273 K at high pressure of 5 GPa, according to our synthesis experiments, as pressure usually favors for the suppression of $N_2$ degassing of TM nitrides. Besides, high-quality single crystals with a suitable crystallite size of 30-40 um can only be prepared above 1673 K at 5 GPa, which is critically important for successful identification of ferroelectricity in $CeTaN_{3-\delta}$. Below 1673 K, the obtained samples have a relatively low crystallinity with a small crystallite size, as manifested by anomalous broadening of diffraction peaks (Fig. S1), which is similar to that observed in $LaWN_{3-\delta}$ [1], hence unfavorable for ferroelectric measurements.

**Table S1. Structural models of $CeTaN_{3-\delta}$ refined by using both XRD and NPD data taken at ambient conditions.** The structural models of $CeTaN_{3-\delta}$ are fetched from ref. [6] and carefully refined based on our diffraction data.

| Structure | Lattice parameter (Å) | Atomic position |
|---|---|---|
| *P*4*mbm* (No.127) | a = 5.6715(5)<br>b = 5.6715(4)<br>c = 3.9931(5) | Ce: 2*d* (0, 1/2, 0)<br>Ta: 2*b* (1/2, 1/2, 1/2)<br>N1: 4*h* (0.829, 0.671, 1/2)<br>N2: 2*a* (1/2, 1/2, 0) |
| *I*4*mcm* (No.140) | a = 5.6717(6)<br>b = 5.6717(6)<br>c = 7.9875(1) | Ce: 4*b* (0, 1/2, 3/4)<br>Ta: 4*c* (1/2, 1/2, 1/2)<br>N1: 8*h* (0.831, 0.331, 1/2)<br>N2: 4*a* (1/2, 1/2, 1/4) |
| *Pnma* (No.62) | a = 5.6870(5)<br>b = 8.0032(3)<br>c = 5.6752(4) | Ce: 4*c* (0.516, 1/4, 0.957)<br>Ta: 4*a* (1/2, 0, 1/2)<br>N1: 8*d* (0.328, 0.027, 0.283)<br>N2: 4*c* (0.010, 1/4, 0.909) |
| $Pmn2_1$ (No.31) | a = 8.0064(2)<br>b = 5.7077(3)<br>c = 5.6825(4) | Ce: 2*a* (1/2, 0.246, 0.013)<br>Ce: 2*b* (1, 0.252, 0.993)<br>Ta: 4*c* (0.257, 0.251, 0.506)<br>N1: 4*c* (0.293, 0.508, 0.289)<br>N2: 4*c* (0.222, 0.010, 0.785)<br>N3: 2*a* (1/2, 0.763, 0.936)<br>N4: 2*b* (1, 0.745, 0.087) |
| $Pna2_1$ (No.33) | a = 5.6754(4)<br>b = 5.6867(6)<br>c = 8.0034(6) | Ce: 4*a* (0.001, 0.997, 0.258)<br>Ta: 4*a* (0.499, 0.001, 0.001)<br>N1: 4*a* (0.278, 0.281, 0.519)<br>N2: 4*a* (0.248, 0.750, 0.455)<br>N3: 4a (0.509, 0.076, 0.257) |
| $Pmc2_1$ (No.26) | a = 8.0061(3)<br>b = 5.6904(5)<br>c = 5.6774(5) | Ce: 2*a* (0, 0.245, 0.033)<br>Ce: 2*b* (1/2, 0.251, 0.037)<br>Ta: 4*c* (0.254, 0.751, 0.018)<br>N1: 4*c* (0.221, 0.972, 0.322)<br>N2: 4*c* (0.285, 0.497, 0.296)<br>N3: 2*a* (0, 0.676, 0.027)<br>N4: 2*b* (1/2, 0.837, 0.037) |
| $Pm\bar{3}m$ (No.221) | a = 4.0116(2) | Ce: 4*a* (1/2, 1/2, 1/2)<br>Ta: 4*a* (0, 0, 0)<br>N: 4*a* (0, 1/2, 0) |

**Table S2. Refined lattice parameters of *Pna*$2_1$-$CeTaN_{3-\delta}$, based on the Rietveld analysis of both NPD and XRD data taken at ambient conditions.**

| | *Pna*$2_1$-$CeTaN_{3-\delta}$ | |
|---|---|---|
| Formula | $CeTaN_{2.66}$ | |
| Symmetry | Orthorhombic, *Pna*$2_1$ (No. 33) | |
| Cell parameter (Å) | a = 5.6754(4), b = 5.6867(3), c = 8.0034(4) | |
| Cell volume (Å$^3$) | 258.309 | |
| Density (g/cm$^3$) | 9.25 | |
| **Atomic position** | **Wyckoff site** | **Occupancy** |
| | Ce: 4*a* (0.001, -0.003, 0.258) | 1 |
| | Ta: 4*a* (0.499, 0.001, 0.001) | 1 |
| | N1: 4*a* (0.278, 0.281, 0.519) | 0.91 |
| | N2: 4*a* (0.248, 0.750, 0.455) | 0.92 |
| | N3: 4a (0.509, 0.076, 0.257) | 0.83 |
| *wRp* (%) | 2.8 for XRD, 6.6 for NPD | |

**Table S3. Refined lattice parameters of *Pmc*2$_1$-CeTaN$_{3-\delta}$ based on the Rietveld analysis of both NPD and XRD data taken at ambient conditions.** In fact, the crystal structure of CeTaN$_{3-\delta}$ can be excellently refined by either *Pna*2$_1$, *Pmc*2$_1$, and *Pmn*2$_1$ with the nearly same excellence of refinements, making them indistinguishable by both the XRD and NPD data. By means of electron diffraction techniques, both *Pna*2$_1$ and *Pmc*2$_1$ can be readily excluded (see Figs. S9 - S11 and Fig. 1 of the main text). As two important candidate models, we also perform the refinements for *Pna*2$_1$- and *Pmc*2$_1$-CeTaN$_{3-\delta}$ and the thus-refined lattice parameters are summarized in Tables S1- S2 for reference.

| ***Pmc*2$_1$-CeTaN$_{3-\delta}$** | | |
|---|---|---|
| Formula | CeTaN$_{2.68}$ | |
| Symmetry | Orthorhombic, *Pmc*2$_1$ (No. 26) | |
| Cell parameter (Å) | *a* = 8.0095 (2), *b* = 5.6779 (4), *c* = 5.6910 (5) | |
| Cell volume (Å$^3$) | 258.810 | |
| Density (g/cm$^3$) | 9.18 | |
| **Atomic position** | **Wyckoff site** | **Occupancy** |
| | Ce: 2*a* (0, 0.245, 0.033) | 1 |
| | Ce: 2*b* (1/2, 0.251, 0.037) | 1 |
| | Ta: 4*c* (0.254, 0.751, 0.018) | 1 |
| | N1: 4*c* (0.221, 0.972, 0.322) | 0.93 |
| | N2: 4*c* (0.285, 0.497, 0.296) | 0.91 |
| | N3: 2*a* (0, 0.662, 0.037) | 0.77 |
| | N4: 2*b* (1/2, 0.818, 0.027) | 0.92 |
| *wRp* (%) | 3.5 for XRD, 6.4 for NPD | |

**Table S4. Refined lattice parameters of *Pmn*$2_1$-$CeTaN_{3-\delta}$ by analysis of NPD data taken at 473, 1073, and 1373 K, respectively.**

| | $CeTaN_{3-\delta}$ | | |
|---|---|---|---|
| Temperature (K) | 473 | 1073 | 1373 |
| Formula | $CeTaN_{2.68}$ | $CeTaN_{2.68}$ | $CeTaN_{2.68}$ |
| Symmetry | Orthorhombic, *Pmn*$2_1$ (No. 31) | Orthorhombic, *Pmn*$2_1$ (No. 31) | Orthorhombic, *Pmn*$2_1$ (No. 31) |
| Cell parameter (Å) | a = 8.0179(3)<br>b = 5.7226(3)<br>c = 5.6887(6) | a = 8.0728(4)<br>b = 5.7538(2)<br>c = 5.7037(4) | a = 8.0919(3)<br>b = 5.7719(4)<br>c = 5.7188(5) |
| Cell volume (Å$^3$) | 260.92 | 264.93 | 269.10 |
| Density (g/cm$^3$) | 9.13 | 8.99 | 8.92 |
| **Atomic position** | | **Wyckoff site** | |
| | Ce: 2*a* (1/2, 0.248, 0.011) | Ce: 2*a* (1/2, 0.257, -0.013) | Ce: 2*a* (1/2, 0.257, 0.001) |
| | Ce: 2*b* (1, 0.254, 0.996) | Ce: 2*b* (1, 0.279, 0.998) | Ce: 2*b* (1, 0.288, 0.986) |
| | Ta: 4*c* (0.248, 0.262, 0.498) | Ta: 4*c* (0.258, 0.264, 0.495) | Ta: 4*c* (0.261, 0.266, 0.491) |
| | N1: 4*c* (0.289, 0.532, 0.285) | N1: 4*c* (0.271, 0.528, 0.272) | N1: 4*c* (0.270, 0.530, 0.264) |
| | N2: 4*c* (0.221, -0.002, 0.767) | N2: 4*c* (0.222, -0.001, 0.754) | N2: 4*c* (0.233, 0.016, 0.754) |
| | N3: 2*a* (1/2, 0.727, 0.935) | N3: 2*a* (1/2, 0.728, 0.926) | N3: 2*a* (1/2, 0.746, 0.928) |
| | N4: 2*b* (1, 0.747, 0.075) | N4: 2*b* (1, 0.750, 0.080) | N4: 2*b* (1, 0.730, 0.064) |

Note: the atomic occupancies are fixed during refinements (see Table 1 of the main text), since the composition of $CeTaN_{3-\delta}$ remains nearly invariant during high-T NPD experiments in vacuum.

## Supplementary references